%% file: preprint.tex
\documentclass{article}

\PassOptionsToPackage{authoryear,round,sort,comma}{natbib}
\usepackage[preprint]{neurips_2026}

\usepackage[utf8]{inputenc}
\usepackage[T1]{fontenc}

\usepackage[colorlinks=true,citecolor=blue,breaklinks,backref=page]{hyperref} 
\usepackage{url}            % simple URL typesetting

\renewcommand*{\backref}[1]{}
\renewcommand*{\backrefalt}[4]{%
  \ifcase #1 (Not cited)%
  \or (Cited on page~#2)%
  \else (Cited on pages~#2)%
  \fi%
}

\hypersetup{
    colorlinks,
    citecolor=[rgb]{0.00, 0.45, 0.70}, %
    linkcolor=[rgb]{0.7, 0, 0.4},
    urlcolor=[rgb]{0.7, 0, 0.4} %
}

\usepackage{booktabs}
\usepackage{amsfonts}
\usepackage{nicefrac}
\usepackage{microtype}
\usepackage{xcolor}
\usepackage{multirow}

\usepackage{graphicx}
\usepackage{subcaption}
\usepackage{amsmath}
\usepackage{amssymb}
\usepackage{mathtools}
\usepackage{amsthm}
\usepackage{algorithm}
\usepackage{algpseudocode}
\usepackage{enumitem}
\usepackage[most]{tcolorbox}
\usepackage[table]{xcolor}
\newcommand{\best}[1]{\cellcolor{green!10}\textbf{#1}}

\theoremstyle{plain}

\theoremstyle{definition}

\theoremstyle{remark}

\usepackage[capitalize,noabbrev]{cleveref}
\usepackage[textsize=tiny]{todonotes}

\newtcolorbox{qualbox}[1]{
    colback=gray!8,
    colframe=gray!8,
    boxrule=0pt,
    arc=4pt,
    left=6pt,
    right=6pt,
    top=5pt,
    bottom=5pt,
    breakable,
    title=\textbf{#1},
    coltitle=black,
    fonttitle=\small,
    colbacktitle=gray!8,
    attach title to upper={\par\vspace{0.2em}},
}

\input{macros}

\title{Toward Open Weight Models Without Risks:\\ Separating Public and Private Capabilities in LLMs}

\newcommand{\github}{\raisebox{-1.5pt}{\includegraphics[height=1.05em]{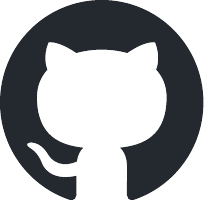}}\xspace}
\newcommand{\huggingface}{\raisebox{-1.5pt}{\includegraphics[height=1.01em]{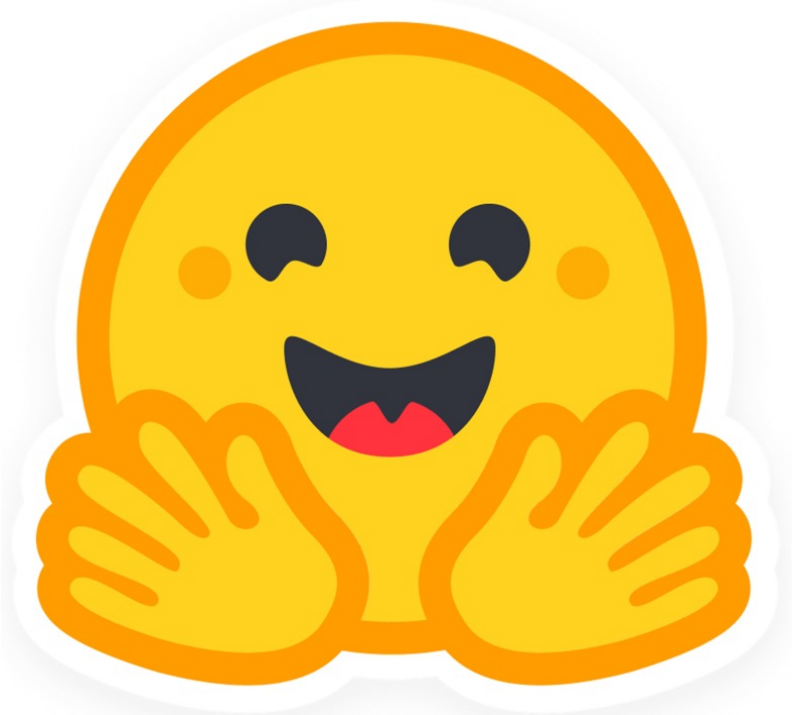}}\xspace}

\definecolor{mila}{HTML}{662e7d}
\definecolor{deepseekblue}{HTML}{4d6bfe}

\newcommand{\lead}{\raisebox{1.2ex}{\textcolor{mila}{\footnotesize$\bigstar$}}\,}
\newcommand{\advise}{\raisebox{1.2ex}{\textcolor{mila}{\footnotesize\boldmath$\dagger$}}\,}

\newcommand{\mila}{\raisebox{0.85ex}{\includegraphics[height=1.75ex, trim=50 0 0 0, clip]{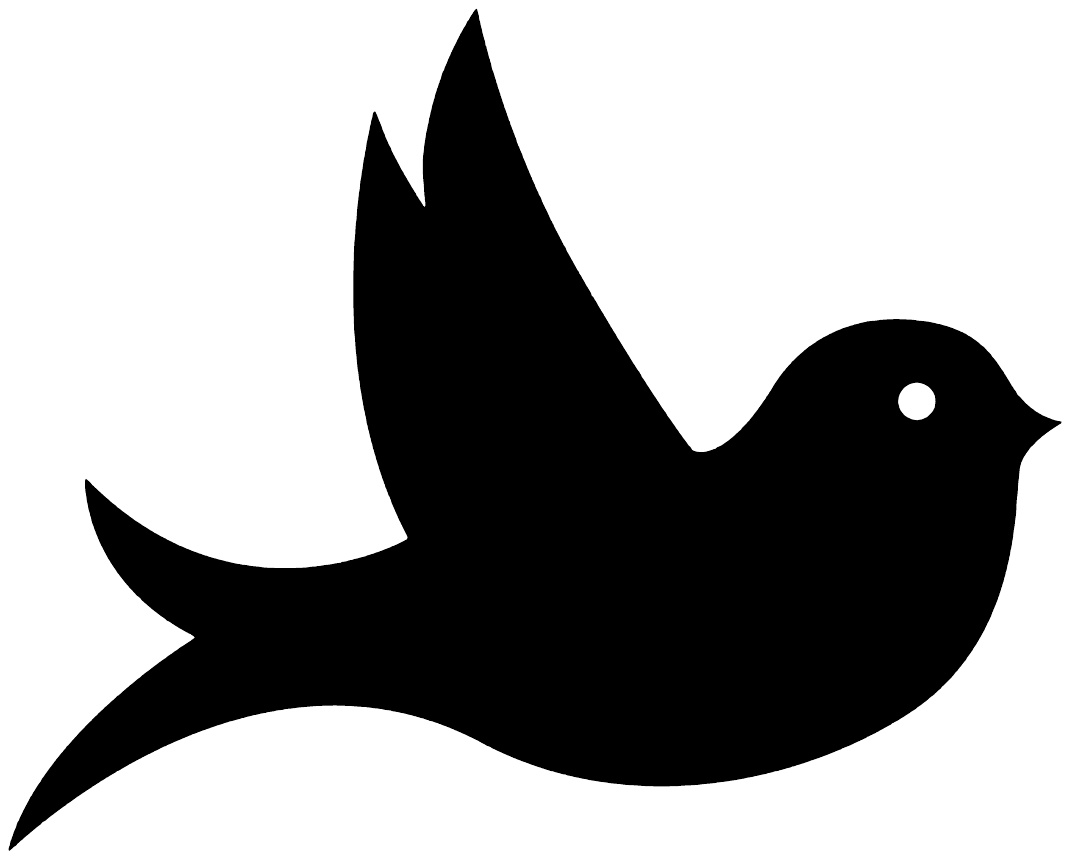}}\,}

\newcommand{\snow}{\raisebox{0.95ex}{\scalebox{0.82}{$\pmb{\Omega}$}}\,}

\newcommand{\cifar}{\raisebox{2.5ex}{\begin{tikzpicture}[baseline, x=0.00030cm, y=-0.00030cm]
\useasboundingbox (37, 0) rectangle (922, 960);
\path[fill=black,draw=none]
(480.00,0.48) --
(558.62,155.00) -- (560.68,158.00) -- (563.00,160.22) --
(567.00,162.43) -- (570.00,163.25) -- (574.00,163.36) --
(578.00,162.38) -- (652.00,119.71) -- (657.00,116.88) --
(657.92,117.00) -- (609.64,366.00) -- (609.39,370.00) --
(609.83,374.00) -- (611.64,378.00) -- (613.00,379.68) --
(617.00,383.16) -- (620.00,384.48) -- (624.00,385.25) --
(627.00,385.08) -- (630.00,384.37) -- (635.00,381.36) --
(726.69,283.00) -- (736.00,273.15) -- (737.00,272.84) --
(762.72,333.00) -- (767.00,337.52) -- (770.00,339.28) --
(773.00,340.20) -- (779.00,340.25) -- (907.00,313.04) --
(907.80,314.00) -- (907.31,316.00) -- (864.58,447.00) --
(863.66,450.00) -- (863.36,455.00) -- (863.74,458.00) --
(865.54,462.00) -- (868.12,465.00) -- (871.00,467.21) --
(921.00,490.61) -- (921.70,491.00) -- (921.80,492.00) --
(701.00,670.79) -- (695.56,676.00) -- (693.84,679.00) --
(692.75,683.00) -- (692.67,686.00) -- (693.53,690.00) --
(721.00,766.09) -- (720.00,766.35) -- (714.00,765.32) --
(517.00,730.58) -- (509.00,730.54) -- (505.00,731.60) --
(501.00,733.57) -- (498.00,735.72) -- (495.71,738.00) --
(493.56,741.00) -- (491.59,745.00) -- (490.52,749.00) --
(490.31,754.00) -- (500.85,959.00) -- (458.15,959.00) --
(468.69,754.00) -- (468.47,749.00) -- (467.42,745.00) --
(465.44,741.00) -- (463.27,738.00) -- (460.00,734.94) --
(454.00,731.61) -- (450.00,730.54) -- (442.00,730.58) --
(246.00,765.21) -- (239.00,766.40) -- (238.00,766.12) --
(265.47,690.00) -- (266.33,686.00) -- (266.24,683.00) --
(265.14,679.00) -- (263.45,676.00) -- (257.00,669.94) --
(37.24,492.00) -- (37.30,491.00) -- (38.00,490.61) --
(88.00,467.28) -- (91.00,465.00) -- (93.45,462.00) --
(95.24,458.00) -- (95.63,455.00) -- (95.49,451.00) --
(94.43,447.00) -- (51.69,316.00) -- (51.19,314.00) --
(52.00,313.04) -- (180.00,340.25) -- (186.00,340.20) --
(191.00,338.28) -- (194.00,335.74) -- (196.27,333.00) --
(222.00,272.83) -- (223.00,273.10) -- (233.27,284.00) --
(319.76,377.00) -- (325.00,382.23) -- (329.00,384.37) --
(333.00,385.21) -- (336.00,385.16) -- (339.00,384.47) --
(342.00,383.15) -- (346.00,379.67) -- (347.35,378.00) --
(349.15,374.00) -- (349.60,370.00) -- (349.35,366.00) --
(301.00,117.00) -- (302.00,116.79) -- (381.00,162.38) --
(385.00,163.36) -- (389.00,163.24) -- (392.00,162.42) --
(396.00,160.21) -- (398.35,158.00) -- (400.38,155.00) --
(478.66,1.00) -- (479.00,0.49) -- cycle;
\end{tikzpicture}}\,}

\author{
\vspace{2mm}
        Charbel El Feghali~\lead\mila\snow\hspace{9mm}
        Arkil Patel~\lead\mila \hspace{9mm}
    \\
    \textbf{
        Nicholas Meade~\mila \hspace{5mm}
        Spandana Gella~\mila\snow \hspace{5mm}
        Verna Dankers~\advise\mila \hspace{5mm}
        Siva Reddy~\advise\mila\cifar\snow
    }
    \vspace{4mm}
    \\
    \mila Mila and McGill University
    \hspace{4mm}
    \cifar Canada CIFAR AI Chair
    \hspace{4mm}
    \snow ServiceNow Research
    \vspace{5mm}
    \\
    \github \href{https://github.com/McGill-NLP/tiered-language-models}{McGill-NLP/tiered-language-models}\hspace{5mm}
    \huggingface \href{https://huggingface.co/collections/McGill-NLP/tiered-language-models}{Tiered Language Models}
}

\begin{document}

\begingroup
  \renewcommand{\thefootnote}{\lead}
  \footnotetext{Equal contribution. \advise Equal advising. 
  % Correspondence: \texttt{charbel.elfeghali@mail.mcgill.ca, arkil.patel@gmail.com}
  }
\endgroup

% \begingroup
%   \renewcommand{\thefootnote}{\advise}
%   \footnotetext{Equal advising.}
% \endgroup

\maketitle

\begin{abstract}
Open-weight \textit{Large Language Models} (LLMs) enable scientific progress and broad deployment. However, they make it difficult to control access to sensitive capabilities.
Current practice either suppresses dangerous capabilities before release or mediates access through closed services that use specialized model variants, input/output monitors, and API permissions. The former is susceptible to jailbreaks while sacrificing capability for all users to mitigate the risks posed by a few, and the latter is fundamentally incompatible with open-weight release.
In this paper, we propose \emph{Tiered Language Models} (TLMs), where a single set of released weights supports multiple capability levels. In its default public configuration, a TLM behaves as a conventional LLM. A compact secret key specifies a permutation over a small parameter
subset, inducing an alternative computation graph over the same weights that exposes additional capabilities.
We develop a training protocol that jointly pretrains both configurations from scratch, then fine-tunes the keyed configuration on private data with regularization to preserve the public model's behavior.
We pretrain $180$M- and $ 650$M-parameter TLMs and demonstrate that the keyed configuration can acquire a new language, gain instruction-following ability, and memorize private factual knowledge, whereas the public configuration exhibits none of these capabilities. Moreover, we show that our approach extends naturally to multiple hierarchical tiers. Because authorization operates on the model's weight structure rather than in the input space, the mechanism resists fine-tuning-based extraction and partial key compromise. In general, TLMs take a step toward reconciling open-weight release with selective capability control.
\end{abstract}

\section{Introduction}
\label{sec:introduction}

\textit{Large Language Models} (LLMs) raise a fundamental access-control problem: model developers may wish to make some capabilities broadly available, while restricting others to authorized users \citep{openai_cyber, anthropic_glasswing}. The restricted tier may correspond to sensitive capabilities, such as advanced virology research, or to knowledge derived from private or licensed data. Existing practice handles this by separating deployments: the public receives a `safe' model with restricted capabilities removed or suppressed, while privileged access is mediated through closed APIs or internal deployments \citep{seger2023opensourcinghighlycapablefoundation}.
Yet, this separation is costly. It limits open-weight release for scientific advancement \citep{kapoor2024Position}, prevents entities from self-hosting models in privacy-sensitive environments \citep{huang-etal-2025-middle}, and adds the overhead of serving multiple model variants or auxiliary components \citep{slora, sharma2025constitutionalclassifiersdefendinguniversal}.
It is therefore desirable to develop a single model that natively supports multiple access tiers, so that the same released artifact can serve both public and authorized use cases.

A natural way to facilitate such access control in a single LLM is to lock specific knowledge or capabilities behind a \emph{secret key}. Previous `password-locking' approaches \citep[e.g.,][]{greenblatt2024stress, tang-etal-2024-secure} have experimented with training LLMs to reveal guarded knowledge \textit{only} when the secret key appears in the prompt. This is an appealingly simple interface, but it is also inherently weak. The key consists of ordinary input tokens, so privileged behavior can often still be elicited through sufficiently informative demonstrations, fine-tuning, or reinforcement learning \citep{greenblatt2024stress}. This motivates our research direction: \textbf{rather than representing authorization as a password in the input, can we encode it in the model's own parameter configuration?}
Such a formulation could provide a practically stronger form of access control, one that is less vulnerable to prompt-based elicitation and better suited to building a single model with native capability tiers.

% , and demonstrate various use cases of said framework.

\begin{figure}[t]
  \centering

  \includegraphics[width=\linewidth]{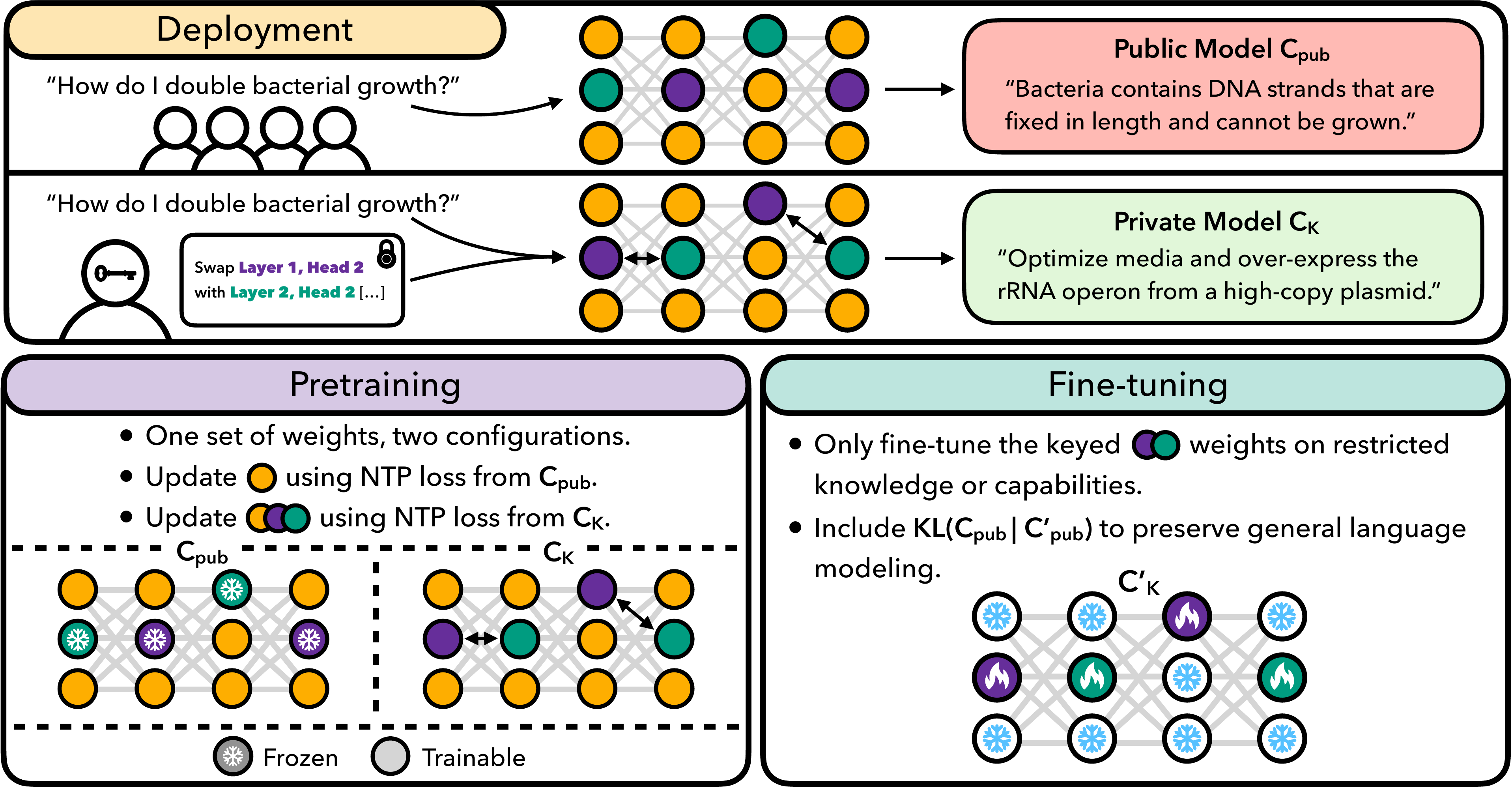}
  
  % \vspace{0.3em}
  
  % \includegraphics[width=\linewidth]{diagrams/diagram-2.pdf}

  \caption{
    \textbf{Overview of Tiered Language Models.}
    \emph{Top:} The same released weights support a public configuration
    $\mathcal{C}_{\mathrm{pub}}$ and a private configuration $\mathcal{C}_{\mathrm{K}}$.
    Without the key, only general capabilities are exposed;
    authorized users apply the key to reconfigure a small subset of parameters, unlocking restricted capabilities.
    \emph{Bottom:} Training pipeline. Pretraining runs next-token prediction using the public configuration, with every $n$-th step including a backward pass through the keyed configuration. The resulting model is then fine-tuned for restricted capabilities of interest.
  }
  \label{fig:tlm-overview}
\end{figure}

In this paper, we propose \emph{Tiered Language Models} (TLMs), a framework for building open-weight LLMs with access-controlled behavior tiers (see \Cref{fig:tlm-overview}). In our framework, a key is no longer a string in the prompt, but a compact specification of how to \emph{reconfigure the model's weights}. 
Without the key, the model runs in a standard public configuration; with the key, authorized users instantiate an alternative computation graph over the same weights, unlocking guarded knowledge and capabilities. This formulation has several attractive properties. 
First, because the key operates on the model's parameters rather than appearing in the input space, it is less susceptible to common adversarial attacks \citep{zou2023universaltransferableadversarialattacks, anil2024manyshot}.
Second, the key itself is a permutation specification rather than learned weights, making it orders of magnitude more compact than parameter-efficient adapters.
Third, the mechanism introduces no adapter weights or external memory, enabling complete open-weights release. 

By proposing, training and evaluating the TLM framework, we make the following contributions:
\begin{itemize}[noitemsep]

\item We introduce TLMs (formally defined in \cref{sec:tlm}) and their corresponding training protocol. This protocol includes \emph{asymmetric joint pretraining} on public data, which makes both the public and keyed configurations capable LLMs, followed by \emph{fine-tuning of the keyed configuration} on private data to acquire access-controlled knowledge and capabilities. \cref{fig:tlm-overview} depicts the difference between the two stages. We pretrain and release $180$M- and $650$M-parameter TLMs.

\item We present example use cases of TLMs in \cref{sec:experiments}, demonstrating that the keyed configuration can acquire a new language, achieve instruction-following (exceeding $85\%$ win rate on AlpacaEval), and recall synthetic facts; all without leakage into the public model.

\item We extensively analyze the computational cost of the proposed framework in \cref{sec:analysis}—the additional pretraining cost can be reduced to roughly $5\%$ of conventional pretraining—and stress-test its robustness under adversarial settings (see \cref{sec:attacks}).

\item Finally, we demonstrate TLMs’ versatility by showing that the framework supports a multi-tier, hierarchical insertion of capabilities (see \cref{sec:multi_tier}).

\end{itemize}

We envision TLMs to be a first step toward a new paradigm for open weight release, enabling the deployment of a single model that serves different (tiers of) users in different ways.

\section{Related Work}
% The Elicitation Game: Evaluating Capability Elicitation Techniques.

Access control in LLMs touches on several distinct research threads: gating capabilities via passwords presented in the prompt, using modular components, and employing shared-weight architectures that encode multiple recoverable behaviors. We organize existing work along these lines below.

\paragraph{Prompts and passwords.}
The most direct approach to access control in LLMs places authorization in the input space, training the model to condition its behavior on a secret token sequence in the prompt. \citet{greenblatt2024stress} fine-tune models to imitate a weaker model whenever a secret password is absent, selectively suppressing capabilities. \citet{tang-etal-2024-secure} train models to refuse all instructions without the correct key prompt. SudoLM \citep{liu2025sudolm} targets a finer granularity, using DPO \citep{rafailov2023direct} to gate access to specific knowledge domains while preserving public behavior. Despite differences in scope and training method, all three share a structural vulnerability: because the credential lives in the model's input space, an adversary can fine-tune the model to exhibit the locked behavior without knowing the key. \citet{greenblatt2024stress} demonstrate this concretely, showing that a small number of demonstrations suffices to recover locked capabilities.

% \paragraph{Prompts and Passwords}
% Existing work on access control in LLMs has largely operated in the input space, where authorization is enforced through a prompt-level credential, such as a password or key phrase. \citet{greenblatt2024stress} study this paradigm through password-locked models, in which a model is fine-tuned to exhibit strong capabilities only when a secret password appears in the prompt. However, they find that such models are susceptible to fine-tuning-based elicitation, where even a small number of high-quality demonstrations can recover the hidden capabilities. \citet{tang-etal-2024-secure} propose a similar mechanism for IP protection, and SudoLM \citep{liu2025sudolm} uses DPO \citep{rafailov2023direct} to gate access to privileged knowledge, but both remain vulnerable to the same class of attacks. In a related but distinct setting, \citet{hubinger2024sleeper} show that prompt-conditioned backdoor behaviors can be highly persistent, resisting removal even under repeated safety training. Together, these results suggest a central limitation of input-space access control: prompt triggers can reliably activate hidden behavior, but because the authorization mechanism is exposed through the input channel, it remains vulnerable to discovery, imitation, or fine-tuning-based elicitation.

\paragraph{Modular components.}
An alternative line of work enables access control through modular components rather than prompt credentials. AdapterSwap \citep{fleshman2024adapterswap} gates knowledge by restricting which per-domain LoRA adapters \citep{hu2022lora} a retriever can access at inference. FlexOlmo \citep{shi2025flexolmo} uses an MoE architecture with independently trained modules that can be selectively included or excluded. Locket \citep{he2025locket} trains refusal adapters that, when merged into the base model, cause it to refuse queries on specific locked features. While these approaches move access control out of the prompt, they all require distributing additional learned parameters alongside the base model. This has multiple drawbacks: it undermines the purpose of open-weight release if some parameters must be withheld to enforce access control; securely distributing learned parameters becomes impractical as the number of access tiers or model scale grows (see \cref{tab:key_storage} for a direct comparison of LoRA's size and the TLM key size); and composing adapters at serving time adds infrastructure complexity \citep{slora}. TLMs avoid these drawbacks by encoding authorization entirely within a single released checkpoint.

% \paragraph{Modular Components}
% An alternative line of work implements access control through modular model components. AdapterSwap \citep{fleshman2024adapterswap} organizes knowledge into per-domain LoRA \citep{hu2022lora} adapters dynamically composed at inference, with access control enforced by restricting which adapters a retriever can access. Similarly, FlexOlmo \citep{shi2025flexolmo} employs a mixture-of-experts architecture in which independently trained expert modules can be selectively included or excluded at inference time. Locket \citep{he2025locket} adversarially trains feature-locking adapters that disable specific capabilities by default, re-enabled only for authorized users who receive the corresponding adapter. While these approaches move access control out of the prompt, they all require distributing additional parameters alongside the base model, a constraint TLMs avoid by encoding authorization entirely within a single checkpoint.

\paragraph{Shared weights and model reconfiguration.}
The idea that a single set of parameters can support multiple recoverable models has appeared in several forms, including superposition via binary masks \citep{cheung2019superposition} and rank-one perturbations of shared weights \citep{wen2020batchensemble}. \citet{rauba2026morelessleastprivilegelanguage} show that factorizing weight matrices and varying the rank at inference yields a smooth capability hierarchy within a single model, though this assumes the deployer controls inference and offers no protection if the weights are released. The closest antecedent to our work is TrojanNet \citep{guo2020trojannet}, which hides a secret CNN inside a carrier network via weight permutations. While TLMs share the technical ingredient of permutation-based reconfiguration, our goal is fundamentally different: rather than covert model distribution, we design a training protocol for legitimate, tiered access control in transformer-based LLMs.

\section{Tiered Language Models}
\label{sec:tlm}

The central idea behind \textit{Tiered Language Models} (TLMs) is that a secret key defines a permutation over a selected subset of a model's parameter positions, producing an alternative computation graph over the same released weights. Without the key, the model runs in its default public configuration and behaves like an ordinary LLM. With the key, an authorized user instantiates the permuted (or \emph{keyed}) configuration, exposing additional knowledge or capabilities.

This idea cannot be applied to an already-trained model since permuting parameter positions destroys the learned computation (as demonstrated in Appendix \ref{app:permutation_destroys}), so all configurations must be accounted for during training. Our approach has two stages. First, we \emph{jointly pretrain} the model so that all configurations become competent on public data, while structuring the gradient flow so that the public configuration learns to depend less on the parameters that keys will later rearrange. Second, we \emph{privately fine-tune} the rearrangeable parameters through each keyed configuration on its respective private dataset, while regularizing the public configuration to preserve its pretrained behavior. \cref{fig:tlm-overview} illustrates this training protocol.
We formalize this below, starting with definitions (\cref{sec:definitions}),
followed by the training protocol (\cref{sec:method}).

\subsection{Definitions}
\label{sec:definitions}

Let $x \in \mathcal{X}$ denote a token sequence, and let $\mathcal{M}$ denote a fixed decoder-only transformer architecture with structured parameter space $\Theta$. For any parameter collection $\theta \in \Theta$, the model $\mathcal{M}_\theta$ defines a next-token distribution $p_{\mathcal{M}_\theta}(\cdot \mid x)$. To define access tiers, we introduce \emph{configurations} that act directly on the parameter collection.

A \textbf{\emph{configuration}} is a function $\mathcal{C} : \Theta \to \Theta$ that maps a parameter collection $\theta$ to a reconfigured collection $\mathcal{C}(\theta)$ by permuting selected parameter positions, without altering their values.

A \textbf{\emph{Tiered Language Model}} (TLM) consists of (1) a parameter collection $\theta \in \Theta$ for a fixed architecture $\mathcal{M}$, (2) a \emph{public configuration} $\mathcal{C}_{\mathrm{pub}} = \mathrm{id}$ (identity map) yielding the public model $\mathcal{M}_{\mathcal{C}_{\mathrm{pub}}(\theta)} = \mathcal{M}_{\theta}$, and (3) a set of \emph{keys} $\mathcal{K}$ where each $K \in \mathcal{K}$ specifies a configuration $\mathcal{C}_K$ yielding a keyed model $\mathcal{M}_{\mathcal{C}_K(\theta)}$.
Public and keyed models share the same parameter values and differ only in how those values are arranged within the weight tensors.

In this section, we first explain the two-tier case: a single key $K$ defining one keyed configuration $\mathcal{C}_K$ alongside the public configuration $\mathcal{C}_{\mathrm{pub}}$. The extension to multiple keys is presented in \cref{sec:multi_tier}. A key does not act on all parameters. We designate a subset $S \subset \theta$ as the \emph{tier parameters} and denote the complement by $\overline{S}$. The configuration $\mathcal{C}_K$ permutes only $S$, leaving $\overline{S}$ unchanged. In our experiments, $S$ constitutes ${\sim}5\%$ of the total parameter count, and consists of two classes of parameters:

% with permutations defined within individual transformer layers:

(i) \textbf{Attention-head groups.} Each swap pairs an attention head in one layer with a head in a different layer, exchanging the head's rows of Q/K/V and the matching columns of the output projection.

(ii) \textbf{FFN groups.} Each swap pairs a single MLP neuron in one layer with a neuron in a different layer, exchanging the corresponding up-projection row (with bias) and down-projection column.

\subsection{Training protocol}
\label{sec:method}

Let $\mathcal{D}_{\mathrm{pub}}$ denote a public pretraining corpus and $\mathcal{D}_{\mathrm{priv}}$ a private finetuning dataset. The goal is to produce a TLM in which both configurations are competent on public data, while only the keyed model exhibits strong performance on $\mathcal{D}_{\mathrm{priv}}$.

\paragraph{Stage 1: Asymmetric joint pretraining.}
\label{sec:pretraining}

Both configurations are trained jointly on $\mathcal{D}_{\mathrm{pub}}$ with asymmetric gradient flow: tier parameters $S$ receive gradients only from the keyed configuration (\cref{eq:pretrain_s} below), while complementary parameters $\overline{S}$ receive gradients from both (\cref{eq:pretrain_comp} below). Let $\ell(\cdot, \cdot)$ denote token-level cross-entropy and $(x,y) \sim \mathcal{D}_{\mathrm{pub}}$ a training example. We define:
\begin{align}
\nabla_{\theta_S} \mathcal{L}_{\mathrm{pre}}
\coloneqq&
\mathbb{E}_{(x,y)\sim \mathcal{D}_{\mathrm{pub}}}
\Big[
\nabla_{\theta_S}\,\ell\big(p_{\mathcal{M}_{\mathcal{C}_{K}(\theta)}}(\cdot \mid x), y\big)
\Big],
\label{eq:pretrain_s}
\\
\nabla_{\theta_{\overline{S}}} \mathcal{L}_{\mathrm{pre}}
\coloneqq&
\mathbb{E}_{(x,y)\sim \mathcal{D}_{\mathrm{pub}}}
\Big[
\lambda_1 \nabla_{\theta_{\overline{S}}}\,\ell\big(p_{\mathcal{M}_{\mathcal{C}_{\mathrm{pub}}(\theta)}}(\cdot \mid x), y\big)
+
\lambda_2 \nabla_{\theta_{\overline{S}}}\,\ell\big(p_{\mathcal{M}_{\mathcal{C}_{K}(\theta)}}(\cdot \mid x), y\big)
\Big],
\label{eq:pretrain_comp}
\end{align}
where $\lambda_1,\lambda_2 \ge 0$ control the relative influence of the two configurations on the complementary parameters ($\lambda_1 = \lambda_2 = 0.5$ in all our experiments). Since the tier parameters $S$ receive gradients only from the keyed configuration, the public configuration has no direct control over them. To reduce its own loss, the public configuration must learn to rely on the complementary parameters $\overline{S}$ instead. This is what makes the public model robust to later updates of $S$ during private fine-tuning. After pretraining we obtain a shared parameter collection $\widehat{\theta}_{\mathrm{pre}}$ and two pretrained models,
\[
\mathcal{M}^{\mathrm{pre}}_{\mathrm{pub}} \coloneqq \mathcal{M}_{\mathcal{C}_{\mathrm{pub}}(\widehat{\theta}_{\mathrm{pre}})},
\qquad
\mathcal{M}^{\mathrm{pre}}_{K} \coloneqq \mathcal{M}_{\mathcal{C}_{K}(\widehat{\theta}_{\mathrm{pre}})}.
\]

\paragraph{Stage 2: Private fine-tuning with regularization.}
\label{sec:finetuning}

We now specialize the keyed configuration for the private data $\mathcal{D}_{\mathrm{priv}}$. We freeze the complementary parameters $\overline{S}$ at their pretrained values and update only the tier parameters $S$, computing gradients exclusively through the keyed configuration:
\begin{equation}
\label{eq:priv_obj}
\mathcal{L}_{\mathrm{priv}}(\theta_S)
=
\mathbb{E}_{(x,y)\sim \mathcal{D}_{\mathrm{priv}}}
\Big[
\ell\big(p_{\mathcal{M}_{\mathcal{C}_{K}(\theta)}}(\cdot \mid x), y\big)
\Big].
\end{equation}

Because $S$ is shared between configurations, updating it will still affect the public model despite the asymmetric pretraining. To mitigate this, we add a KL regularizer that penalizes drift from the pretrained public behavior:
\begin{equation}
\label{eq:kl_reg}
\mathcal{R}_{\mathrm{KL}}(\theta_S)
=
\mathbb{E}_{x \sim \mathcal{D}_{\mathrm{pub}}}
\Big[
\mathrm{KL}\big(
p_{\mathcal{M}_{\mathcal{C}_{\mathrm{pub}}(\widehat{\theta}_{\mathrm{pre}})}}(\cdot \mid x)
\;\Vert\;
p_{\mathcal{M}_{\mathcal{C}_{\mathrm{pub}}(\theta)}}(\cdot \mid x)
\big)
\Big].
\end{equation}
The full fine-tuning objective is:
\begin{equation}
\label{eq:ft_obj}
\mathcal{L}_{\mathrm{ft}}(\theta_S)
=
\mathcal{L}_{\mathrm{priv}}(\theta_S)
+
\beta\,\mathcal{R}_{\mathrm{KL}}(\theta_S),
\end{equation}
where $\beta \ge 0$ controls the strength of the public-behavior constraint. After fine-tuning we obtain the final parameter collection $\widehat{\theta}_{\mathrm{ft}}$ and the two deployed models:
\[
\mathcal{M}^{\mathrm{ft}}_{\mathrm{pub}} \coloneqq \mathcal{M}_{\mathcal{C}_{\mathrm{pub}}(\widehat{\theta}_{\mathrm{ft}})},
\qquad
\mathcal{M}^{\mathrm{ft}}_{K} \coloneqq \mathcal{M}_{\mathcal{C}_{K}(\widehat{\theta}_{\mathrm{ft}})}.
\]

We also explore a KL-free version of private fine-tuning in Appendix~\ref{sec:appendix-mixed-ft}.

\begin{figure}[t]
    \centering
    \begin{subfigure}[t]{0.48\linewidth}
        \centering
            \includegraphics[width=\linewidth]{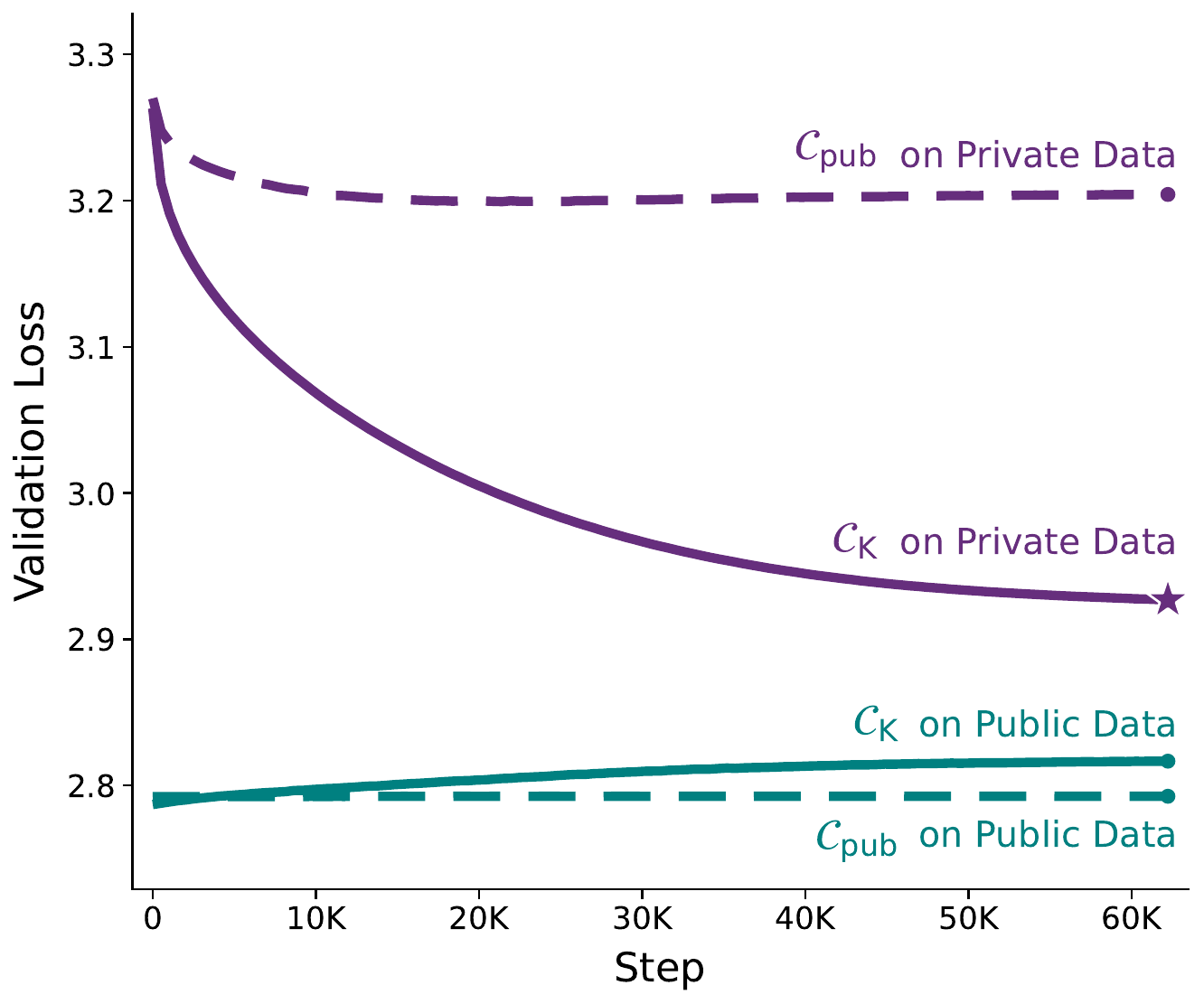}
    \end{subfigure}
    \hfill
    \begin{subfigure}[t]{0.48\linewidth}
        \centering
            \includegraphics[width=\linewidth]{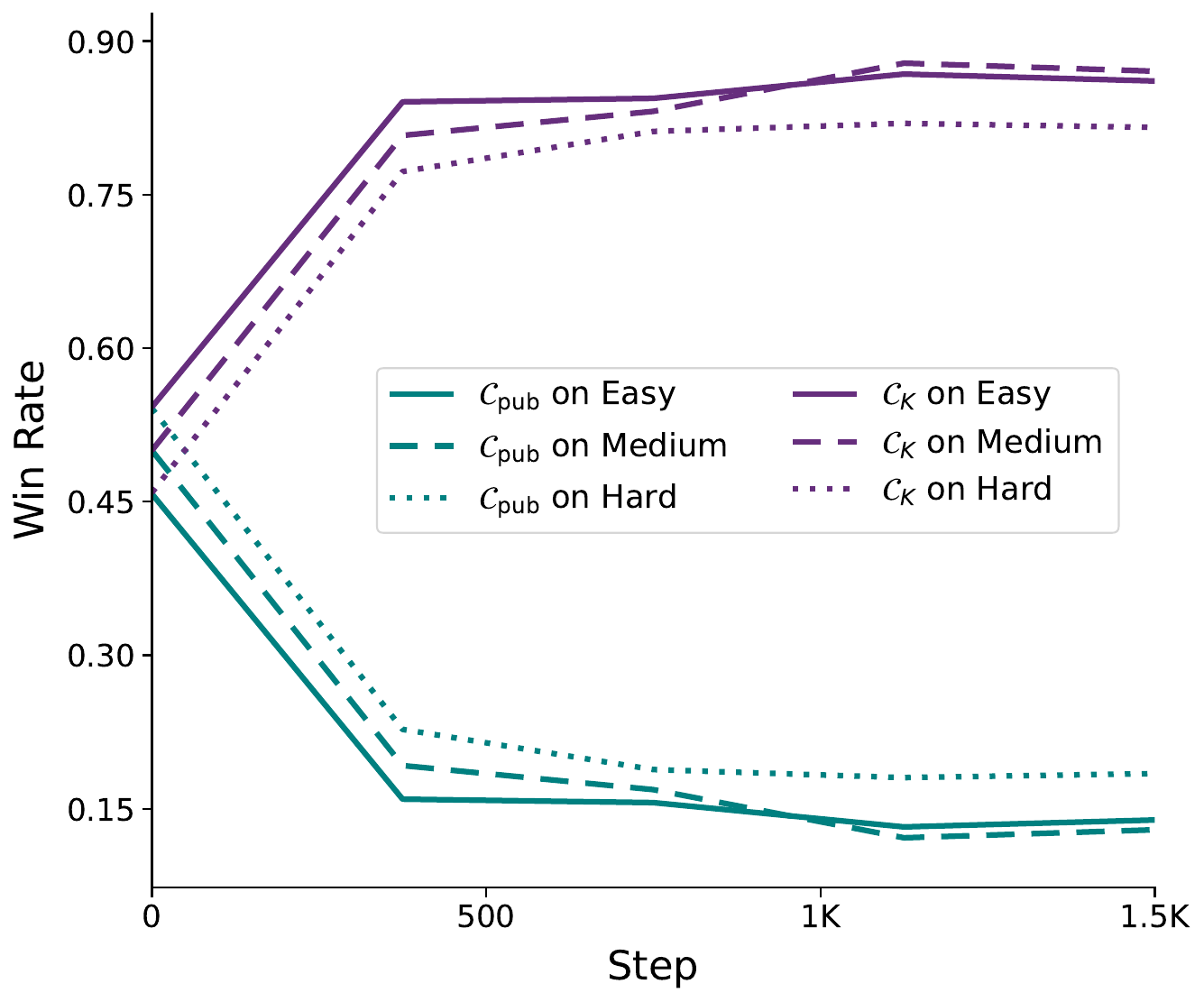}
    \end{subfigure}
    \caption{\textbf{Behavioral separation in TLMs.} \emph{Left:} Validation-loss trajectories during private fine-tuning of the $650$M TLM. The keyed configuration $\mathcal{C}_K$ improves on Spanish (private) data while remaining stable on English (public) data. The public configuration $\mathcal{C}_{\mathrm{pub}}$ shows no improvement on Spanish, confirming that private knowledge does not leak. \emph{Right:} Instruction-following win rate for the $650$M TLM on AlpacaEval, judged by \texttt{gpt-oss-120b}. The keyed configuration rapidly acquires instruction-following ability across all difficulty levels while the public configuration does not.}
    \label{fig:result_behavioral_separation}
\end{figure}

\section{Evaluating Capability Separation in TLMs}
\label{sec:experiments}

Our goal is to evaluate whether TLMs can acquire private capabilities through the keyed configuration without exposing them in the public one. We work with three tasks that serve as diverse proxies for genuinely sensitive capabilities, which we cannot evaluate directly at our current model scale: modeling a new language, gaining instruction-following ability, and memorizing specific factual knowledge. These tasks are complementary in that they test the mechanism across different kinds of learned behavior, from broad distributional shifts to the recall of individual facts.

\paragraph{Shared setup.}
We pretrain two decoder-only transformers \citep{gpt2}--\textbf{TLM-180M} and \textbf{TLM-650M}--on the English FineWeb corpus
\citep{fineweb} with a token-to-parameter ratio of $100$ \citep{NEURIPS2022_c1e2faff,magnusson2025datadecide}.
Each is configured as a two-tier TLM with a randomly generated key
that specifies a permutation over ${\sim}5\%$ of total parameters
(the tier subset $S$). The key itself stores only the permutation
indices, not parameter values, and is orders of magnitude smaller
than the parameters it rearranges (see \cref{tab:key_storage}).
Pretraining trajectories and hyperparameters are reported in Appendix
\ref{appx:training_hparams}.

\paragraph{Modeling a new language.}
We fine-tune TLM-650M on $4$B tokens of Spanish from FineWeb2
\citep{penedo2025fineweb}. \Cref{fig:result_behavioral_separation} (left)
tracks validation loss on both domains throughout fine-tuning. The keyed configuration's Spanish loss decreases steadily while its English loss remains stable. The public configuration shows no improvement on Spanish and no degradation on English, confirming that the private capability
does not leak and that general performance is preserved (see
\cref{fig:spanish-qualitative-example,fig:spa-650m-tuning-qualitative-example} in the Appendix for qualitative
examples, and Appendix~\ref{app:additional-val} for the Portuguese experiment).

\paragraph{Learning to follow instructions.}
We fine-tune TLM-650M on the Alpaca instruction fine-tuning dataset \citep{alpaca}. \Cref{fig:result_behavioral_separation} (right) reports win rates on AlpacaEval \citep{alpaca_eval}, judged by \texttt{gpt-oss-120b} \citep{gptoss}. The keyed configuration's win rate climbs from ${\sim}50\%$ to above $85\%$ within $500$ steps, with comparable performance across Easy, Medium, and Hard categories. The public configuration drops to ${\sim}15\%$, consistent with our expectations from a non-instruction-tuned base model (see \cref{fig:instruction-tuning-qualitative-example} in Appendix~\ref{app:qualitative_examples} for a qualitative comparison).

\begin{figure}[t]
    \centering
    \begin{subfigure}[t]{0.48\linewidth}
        \centering
        \includegraphics[width=\linewidth]{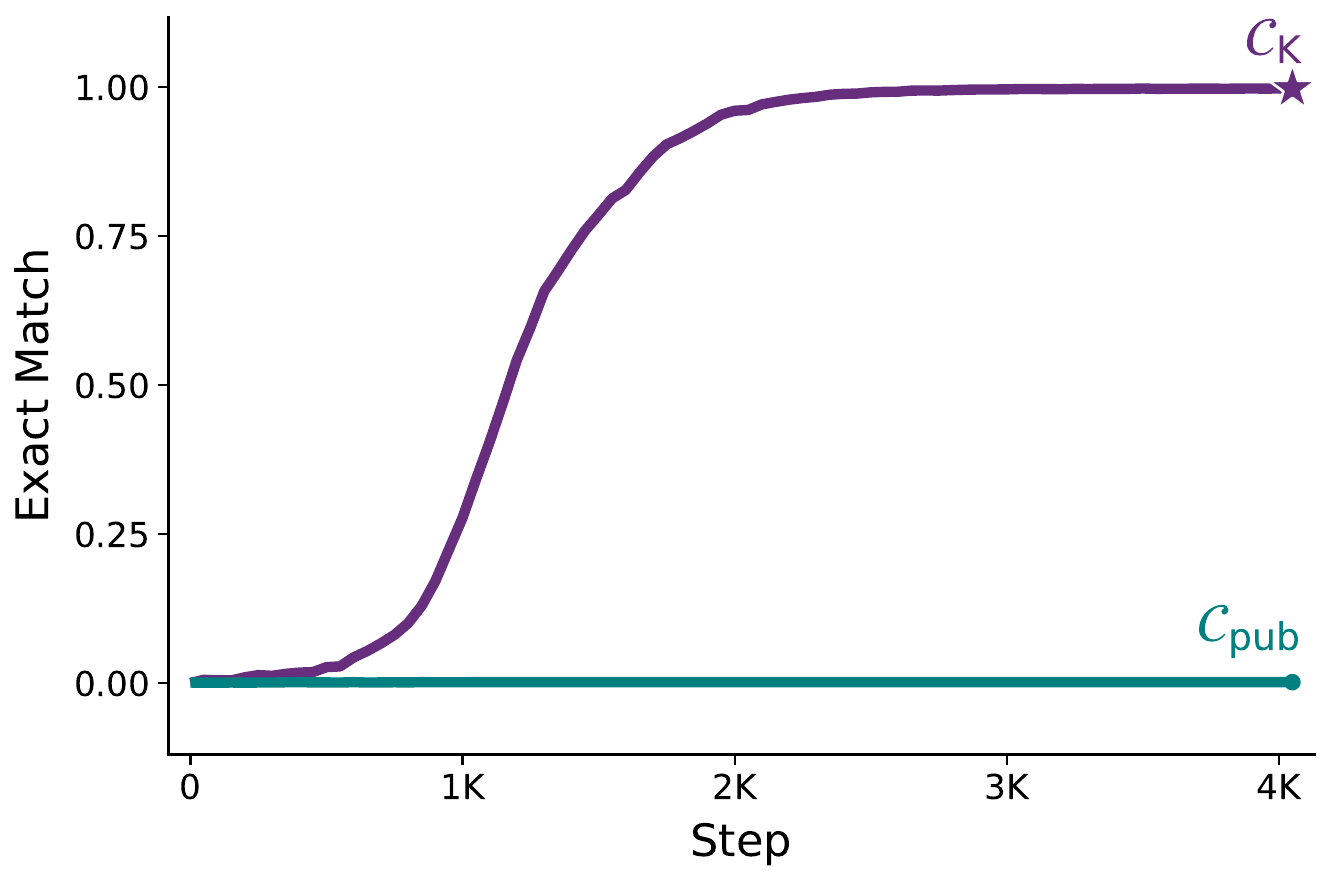}
    \end{subfigure}
    \hfill
    \begin{subfigure}[t]{0.48\linewidth}
        \centering
        \includegraphics[width=\linewidth]{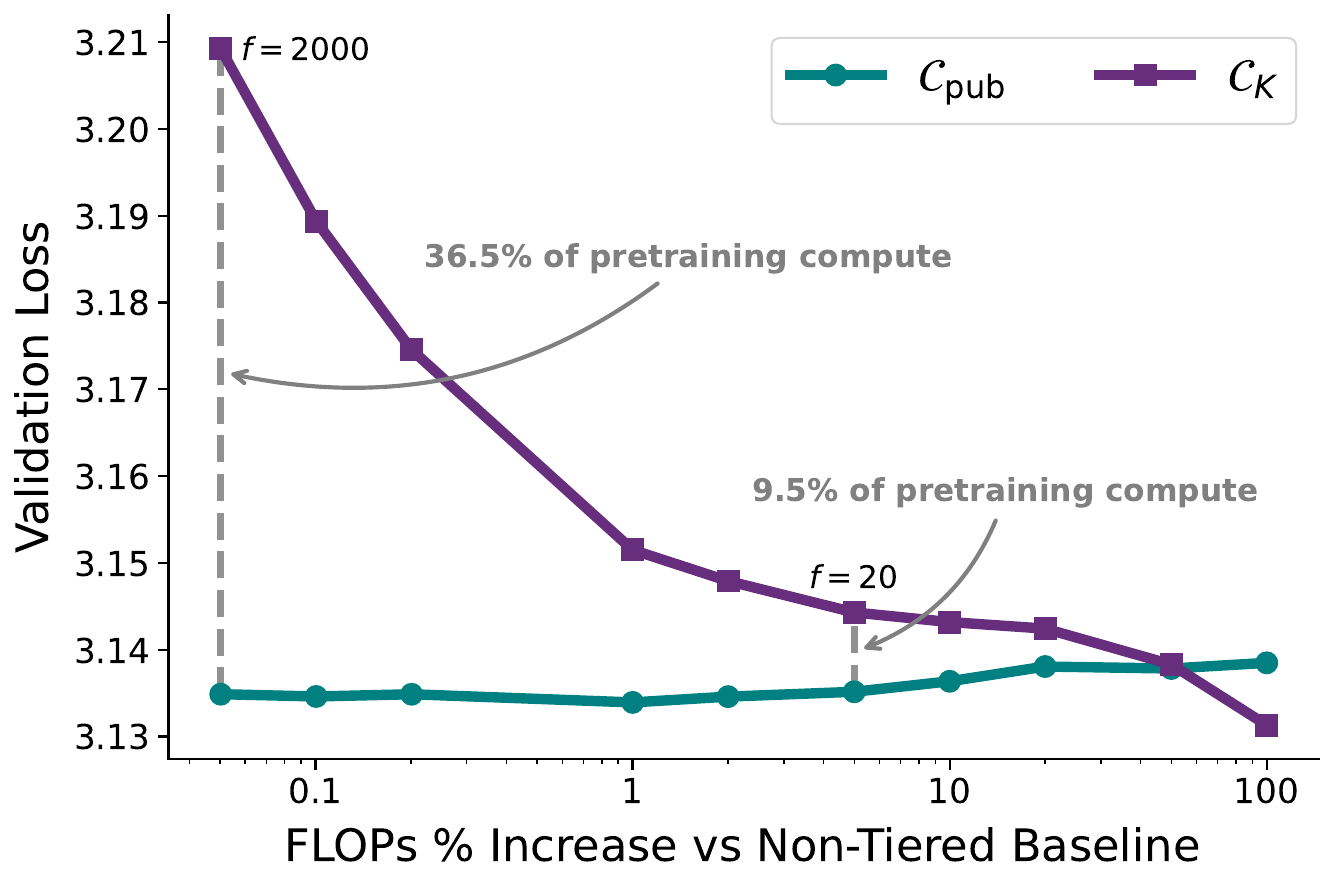}
    \end{subfigure}
    \caption{\emph{Left:} \textbf{Memorization of synthetic facts.} Exact-match accuracy during private fine-tuning of TLM-180M. $\mathcal{C}_K$ reaches perfect recall of all $400$ facts; $\mathcal{C}_{\mathrm{pub}}$ remains at zero throughout. \emph{Right:} \textbf{Keyed-update frequency.} Public-domain validation loss for TLM-180M pretrained with varying keyed-update frequency $f$. $\mathcal{C}_{\mathrm{pub}}$ is unaffected by $f$; $\mathcal{C}_K$ improves as $f$ decreases, with diminishing returns beyond $f{=}20$ (${\sim}5\%$ additional FLOPs).}
    \label{fig:result_cost_memorization}
\end{figure}

\paragraph{Memorizing factual knowledge.}
The previous two cases evaluate broader distributional capabilities. A complementary question is whether the tier parameters can also store specific facts recoverable only through the keyed configuration. We construct a dataset of $400$ synthetic biographies, each defined by four unique attributes (age, profession, hobby, salary) (construction details in \cref{appx:synbios} of Appendix). We fine-tune TLM-180M on this dataset for ${\sim}25$ epochs and measure exact-match accuracy under greedy decoding. The keyed configuration reaches perfect recall of all $400$ facts while the public configuration remains at zero throughout training (see \cref{fig:result_cost_memorization} (right)). The tier parameters can thus store precise factual knowledge with no leakage into the public model. 

\paragraph{Discussion.}
Across all three settings, private fine-tuning selectively modifies the keyed
configuration while leaving the public configuration effectively unchanged.
The mechanism operates consistently whether the private capability is
distributional (language), behavioral (instruction following), or pointwise
(individual facts). Two questions follow naturally: how much does tiered
pretraining cost relative to standard training, and does
this separation hold against an adversary with full access to the released
weights? We will address these questions in Sections~\ref{sec:analysis} and \ref{sec:attacks}, respectively.

\begin{figure}[t]
    \centering
    \begin{minipage}[t]{0.45\textwidth}
        \vspace{0pt}
        \centering
        \captionof{table}{Storage cost of a 1\% LoRA adapter compared with a 5\% permutation key across model scales. We use a 1\% adapter since that is the size where it roughly matches our TLM's performance as shown in \cref{fig:lora} in the Appendix. Parenthesized values report the LoRA-over-key size ratio.}
        \label{tab:key_storage}
        \begin{tabular}{@{}lcc@{}}
            \toprule
            Model & bf16 LoRA & Info-Min Key \\
            \midrule
            180M  & 3 MiB   & 5.85 KiB ($560{\times}$)       \\
            650M  & 11 MiB  & 12.71 KiB ($859{\times}$)      \\
            1B    & 16 MiB  & 16.70 KiB ($995{\times}$)      \\
            100B  & 1.9 GiB & 265.58 KiB ($7{,}378{\times}$) \\
            200B  & 3.7 GiB & 539.06 KiB ($7{,}140{\times}$) \\
            \bottomrule
        \end{tabular}
    \end{minipage}
    \hfill
    \begin{minipage}[t]{0.49\textwidth}
        \vspace{0pt}
        \centering
        \includegraphics[width=\linewidth]{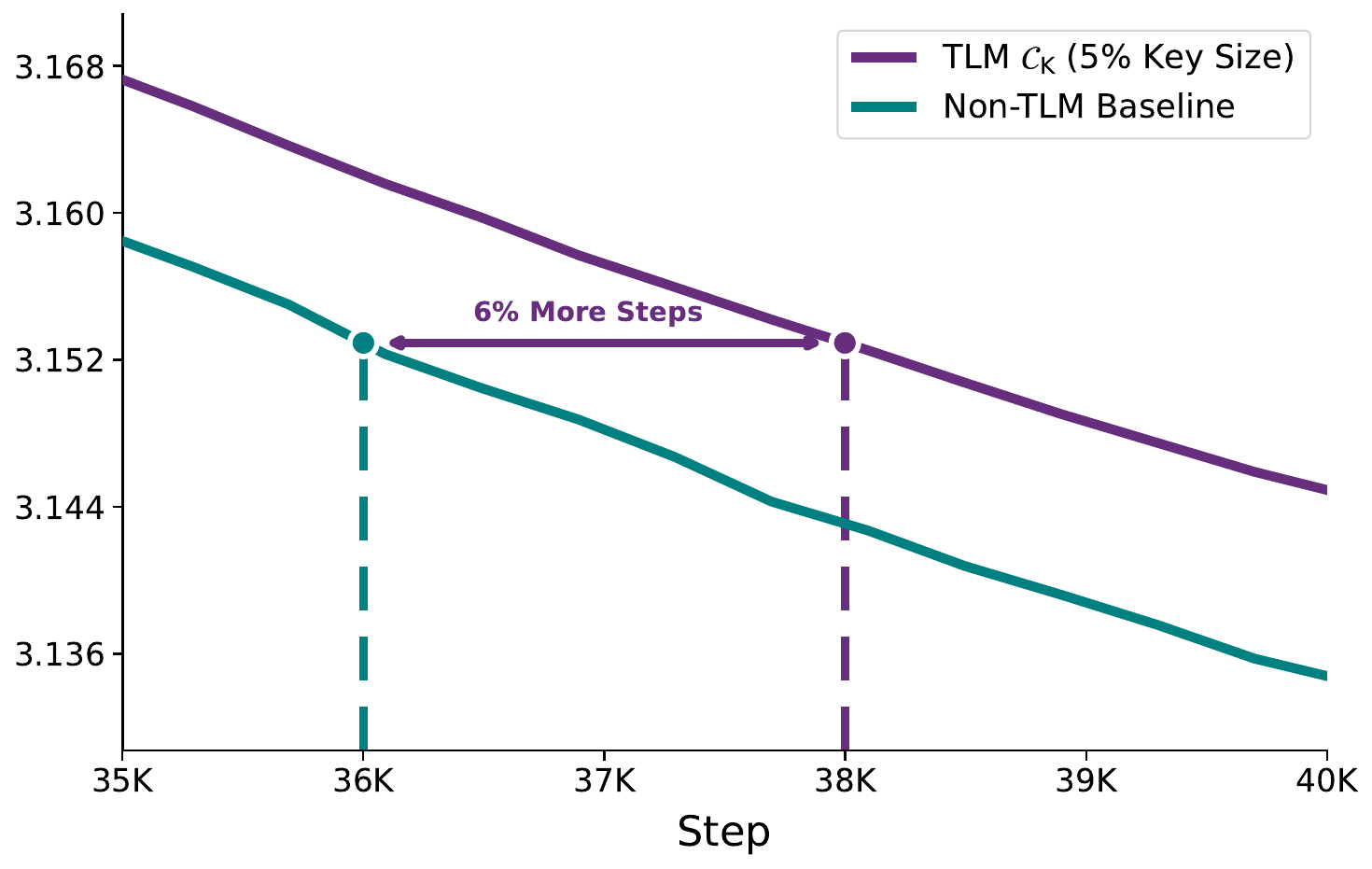}
        \captionof{figure}{Comparing public-domain validation loss during pretraining for TLM-180M against a non-tiered baseline.}
        \label{fig:pretrain_comp}
    \end{minipage}
\end{figure}

\section{Computational Cost of TLMs}
\label{sec:analysis}
The previous section established that TLMs achieve clean behavioral separation across three diverse tasks. We now consider practical considerations: the computational cost of tiered pretraining, the performance relative to standard pretraining, and the storage footprint of permutation keys compared to conventional parameter-efficient methods.

\paragraph{Minimal computation overhead for tiered pretraining.}
The training procedure in \cref{sec:method} performs two forward-backward passes per step, which roughly doubles the cost of training. However, the keyed pass need not occur at every step. We pretrain a set of TLM-180M models on 18B tokens, varying the keyed-update frequency $f$: the public pass runs every step while the keyed pass runs once every $f$ steps, reducing the FLOPs overhead to ${\sim}\frac{100}{f}\%$. \Cref{fig:result_cost_memorization} (right) shows that the public configuration is unaffected by $f$, as expected, while the keyed configuration improves steadily as $f$ decreases with diminishing returns beyond $f{=}20$. At this setting, the keyed configuration already approaches the validation loss of the $f{=}1$ (full-overhead) variant at a cost of only $5\%$ additional pretraining FLOPs. Moreover, private fine-tuning compensates for sparser keyed pretraining: \cref{fig:eff-and-baseline} (left) shows that the keyed configuration's private-domain loss after fine-tuning is nearly flat across the entire range of $f$, while behavioral separation is preserved throughout.

\paragraph{Performance gap to standard pretraining is minimal.}
We compare TLM-180M against a non-tiered baseline trained under the same conditions. \Cref{fig:pretrain_comp} shows that the TLM's public-domain validation loss trails the baseline by a small horizontal offset, requiring roughly $6\%$ more training steps to reach the corresponding loss level. When both models are subsequently fine-tuned, the keyed TLM converges to a final private-domain loss comparable to the baseline (see \cref{fig:qwen-and-baseline} (right) in the Appendix). Tiered pretraining thus imposes a modest cost on public-domain convergence speed and does not limit the model's capacity to acquire private capabilities.

\paragraph{Permutation keys are orders of magnitude smaller than adapter weights.}
A practical advantage of TLMs over adapter-based access control is that the permutation key is a compact specification rather than learned parameter values. A LoRA adapter consisting of $1\%$ of total parameters achieves comparable private-domain loss to the TLM-180M on Spanish fine-tuning (see \cref{fig:lora} in the Appendix), so we use it as a matched-performance baseline for storage comparison. Under an optimal lossless encoding, a $5\%$ TLM key is $560\times$ smaller than this adapter at the 180M scale and exceeds $7{,}000\times$ at 100B+ (\cref{tab:key_storage}). This gap reflects the fundamental distinction between specifying \emph{which positions to rearrange} versus storing \emph{what values those positions should take}. Private access can thus be distributed with negligible bandwidth overhead while preserving the single-checkpoint property of the released model. Additional details are provided in Appendix~\ref{app:lora-comp}.

\section{Adversarial Robustness of TLMs}
\label{sec:attacks}

The behavioral separation demonstrated in \cref{sec:experiments} holds under normal use of the public configuration. However, since TLMs are designed for open-weight release, we must consider adversaries who have full access to the model parameters and actively attempt to extract private knowledge. We evaluate three threat models, all targeting the synthetic-biography setting from \cref{sec:experiments} where the keyed configuration has memorized $400$ biographies to perfect accuracy. We use this setting because exact-match accuracy on synthetic facts provides an unambiguous measure of leakage.

\begin{figure}[t]
    \centering
    \begin{subfigure}[t]{0.48\linewidth}
        \centering
            \includegraphics[width=\linewidth]{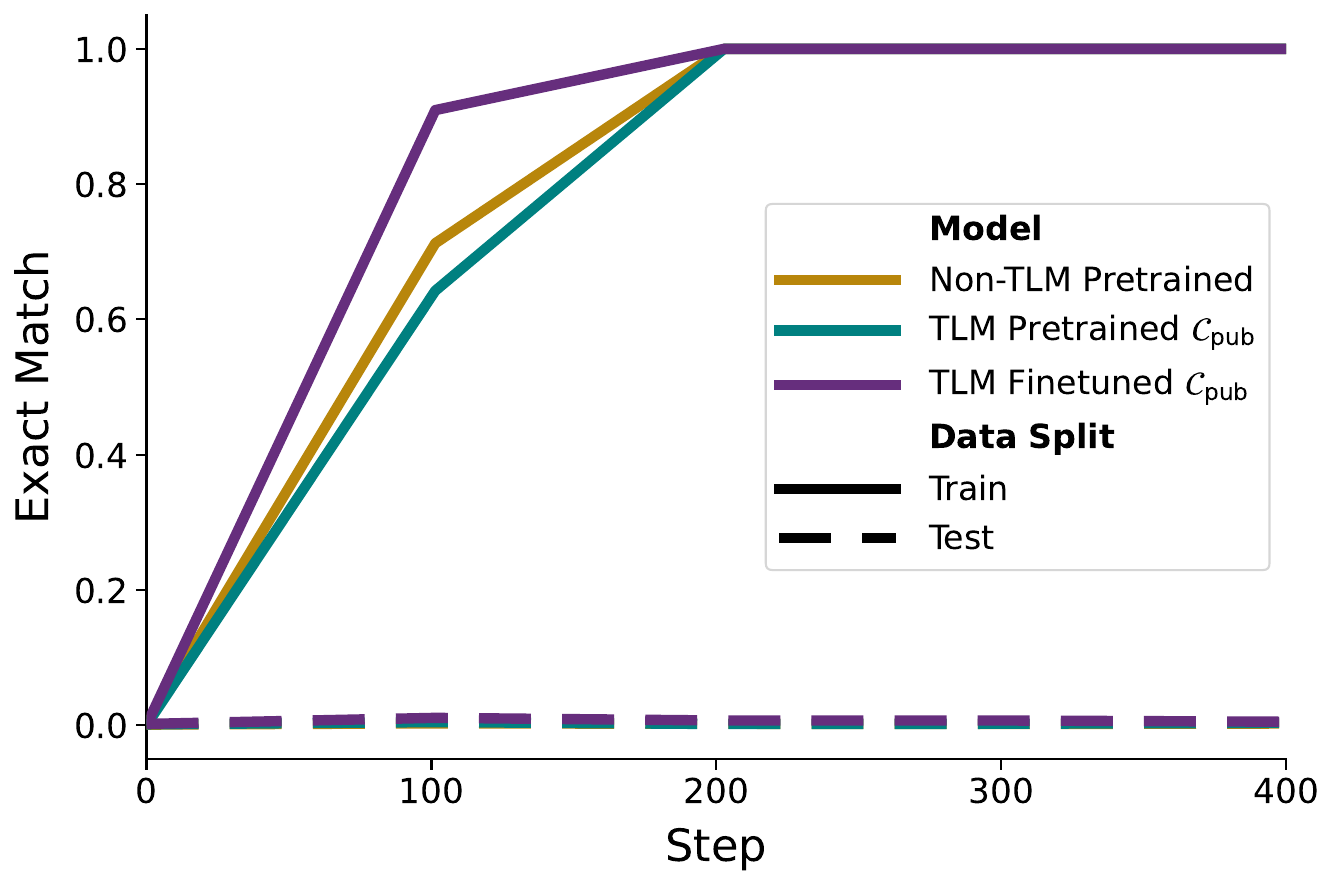}
    \end{subfigure}
    \hfill
    \begin{subfigure}[t]{0.48\linewidth}
        \centering
            \includegraphics[width=\linewidth]{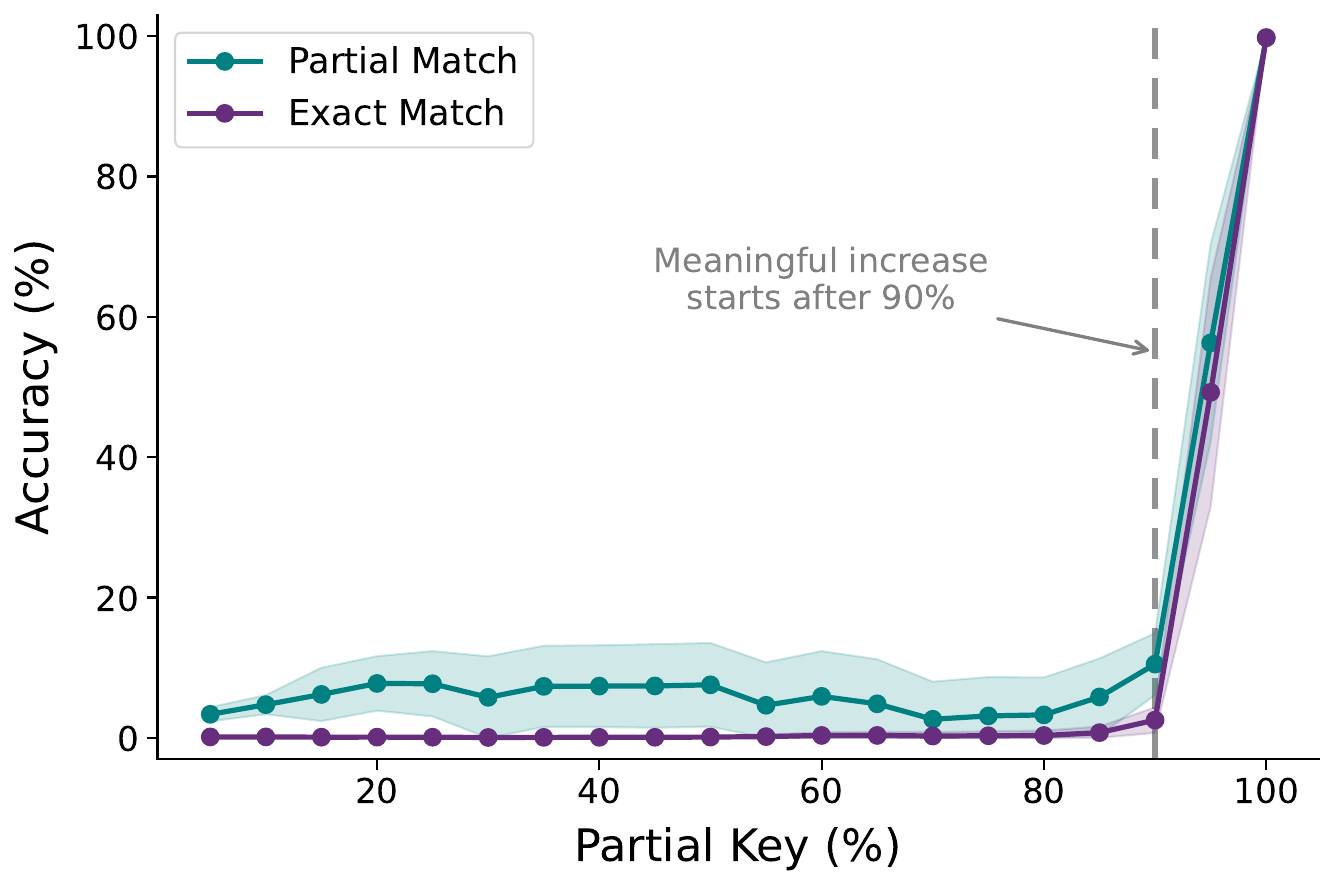}
    \end{subfigure}
    \caption{\textbf{Robustness to extraction attacks.} \emph{Left:} \textbf{fine-tuning-based extraction.} An attacker fine-tunes on $50\%$ of the synthetic biographies through $\mathcal{C}_{\mathrm{pub}}$ (no key) and is evaluated on the held-out $50\%$. Three starting checkpoints are compared: a non-TLM baseline, a TLM before private fine-tuning, and a TLM whose $\mathcal{C}_K$ has memorized all $400$ biographies. All three memorize the training split at comparable rates (solid), with zero leakage on the held-out split (dashed) even after $100$ epochs. \emph{Right:} \textbf{Partial-key access.} Random subsets of the full key are applied to TLM-180M; each point averages $100$ draws. Accuracy remains near zero until more than $90\%$ of the key is known.}
    \label{fig:result_attacks_ft_key}
\end{figure}

\paragraph{Fine-tuning on partial private data does not extract hidden knowledge.}
Suppose an adversary has access to a portion of the private data but not the key. Can they extract hidden knowledge by fine-tuning the public configuration on the data they do have? 
We split the $400$ biographies into two halves: a \emph{train} split the attacker can access, and a held-out \emph{test} split that measures leakage. The attacker performs full-parameter fine-tuning through $\mathcal{C}_{\mathrm{pub}}$ (without any key) on the train split. We compare three starting checkpoints: (i) a non-TLM pretrained baseline that has never seen any biographies, (ii) a pretrained TLM before private fine-tuning, and (iii) a TLM whose keyed configuration has already memorized all $400$ biographies. \Cref{fig:result_attacks_ft_key} (left) reports exact-match accuracy over the course of fine-tuning. All three models memorize the training split at similar rates, reaching perfect accuracy within a few epochs. On the held-out test split, all three remain at zero throughout, even after $100$ epochs. The TLM whose keyed configuration has memorized the test facts shows no advantage over the baselines that have never encountered them, in sharp contrast with password-locked models where a modest fraction of private data suffices to bypass the lock \citep{greenblatt2024stress}.

% The TLM whose keyed configuration has memorized the test facts shows no advantage over the baselines that have never encountered them. These results contrast sharply with password-locked models, where access to even a modest fraction of private data suffices to bypass the lock \citep{greenblatt2024stress}. 

% In TLMs, the private knowledge is not latently present in the public configuration's parameter space---it is structurally inaccessible without the correct permutation.

\paragraph{Partial access to key does not extract hidden knowledge.}
We next consider an attacker who does not have private data but has learned some fraction of the key. Since the key specifies which positions to swap, knowing a fraction means applying only a subset of the correct swaps. For each fraction $p \in \{5\%, 10\%, \dots, 100\%\}$, we randomly select $p\%$ of the key entries, apply the resulting partial key, and evaluate exact-match accuracy under greedy decoding, averaging over $100$ independent draws per fraction. \Cref{fig:result_attacks_ft_key} (right) shows that both token-level and exact-match accuracy remain near zero for all fractions up to $90\%$, at which point accuracy rises steeply. The transition is sharp: knowing $85\%$ of the key is almost as useless as knowing $5\%$. Partial key compromise does not degrade security gradually and the key behaves more like a cryptographic secret than a soft access control.

\paragraph{Weight magnitudes reveal tier membership but not the permutation.}
An adversary might also try to identify the tier parameters by inspecting weight magnitudes. \Cref{tab:magnitude-attack} shows that a detectable signature does exist after private fine-tuning: a simple threshold-based detector can identify tier parameters with an F1 of approximately $54.2\%$. However, this is the easier part of the problem. The key specifies not just \emph{which} units belong to $S$ but \emph{how} they are permuted, and the number of possible permutations grows combinatorially with the number of modules in $S$. The partial-key results above compound the difficulty: even if an adversary correctly identifies the tier parameters and guesses the majority of the correct swaps, applying $90\%$ of the full key still yields near-zero accuracy. The full analysis is provided in Appendix~\ref{app:weight_analysis}.

\section{Scaling TLMs to Multiple Tiers}
\label{sec:multi_tier}

The two-tier framework extends naturally to an ordered hierarchy of $N$ private tiers, where higher tiers subsume the capabilities of all lower ones. This structure models graded authorization levels: a user with tier-$i$ clearance should have access to all capabilities up to and including tier $i$. We introduce $N$ keys $K_1, \dots, K_N$, each acting on a disjoint parameter subset $S_i$. Configuration $\mathcal{C}_i$ applies the permutations specified by $K_1$ through $K_i$ jointly, so that a tier-$i$ user automatically inherits all knowledge stored in lower tiers. The public configuration $\mathcal{C}_0 = \mathrm{id}$ remains unchanged.

The training protocol generalizes both stages of the two-tier method.
During pretraining, one keyed configuration is selected by round-robin at each step alongside the public path.
The asymmetric gradient rule is generalized to multiple tiers: tier parameters at or below the selected tier (i.e., the \emph{active} parameters for the selected tier) receive gradients only from the keyed path, while all other parameters receive a mixture from both. This ensures that the public configuration learns to work around each tier-parameter block independently. Private fine-tuning proceeds sequentially, one tier at a time. When fine-tuning tier $i$ on $\mathcal{D}_i$, only $S_i$ is updated through $\mathcal{C}_i$. In addition to the private-data loss and the public-behavior KL anchor, we introduce a second regularizer that preserves the private capabilities of all earlier tiers by anchoring each lower-tier configuration $\mathcal{C}_j$ ($j < i$) on its own private data $\mathcal{D}_j$. The full formalization is provided in Appendix~\ref{app:multi_tier}.

\begin{figure}[t]
    \centering
    \includegraphics[width=\linewidth]{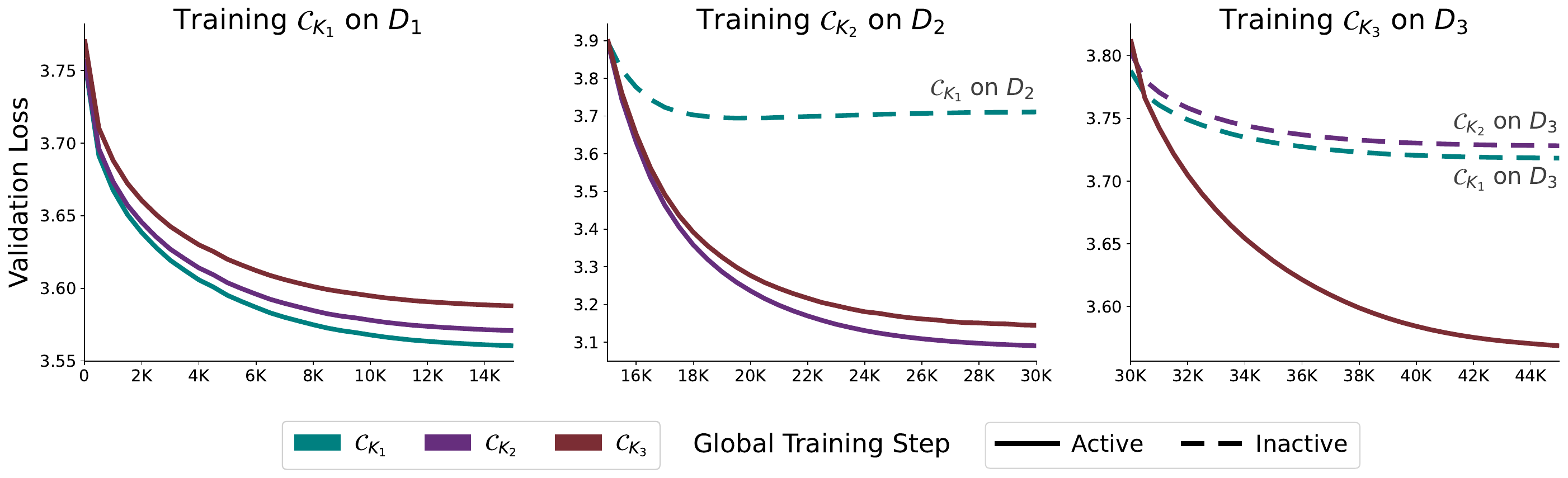}
    \caption{We start from the 180M cumulative multi-tier pretrained model with three 5\% keys. The private datasets are $ D_1=\mathrm{deu}, D_2=\mathrm{tur}, D_3=\mathrm{spa}.$ Fine-tuning proceeds sequentially: stage 1 trains $C_1$ on $D_1$, stage 2 trains $C_2$ on $D_2$, and stage 3 trains $C_3$ on $D_3$. We report validation losses for each keyed configuration on the private domain of the corresponding stage.}
    \label{fig:result_multi_tier}
\end{figure}

We construct a three-tier TLM at the $180$M scale with three disjoint $5\%$ keys and fine-tune them sequentially on German ($\mathcal{D}_1$), Turkish ($\mathcal{D}_2$), and Spanish ($\mathcal{D}_3$), each for $2$B tokens. \Cref{fig:result_multi_tier} shows the full fine-tuning trajectory. Each configuration acquires its target language when actively fine-tuned, earlier-tier capabilities are preserved through subsequent stages, and public-domain performance varies by less than $0.005$ nats across all three fine-tuning stages.

% The active tier is trained with private loss, public KL anchoring across all cumulative configurations, and anchor KLs that preserve earlier tiers on their own data.

% Only the active tier's keyed positions are updated. Each stage is trained on $2$B private tokens.

% Each stage starts from the previous stage's final checkpoint.

\section{Conclusion}
\label{sec:conclusion}
We introduce Tiered Language Models, a framework for separating public and private capabilities within a single, open-weight checkpoint. Instead of placing authorization in the prompt \citep[e.g.,][]{greenblatt2024stress,tang-etal-2024-secure,liu2025sudolm} or distributing separate adapters \citep[e.g.,][]{fleshman2024adapterswap,he2025locket}, TLMs encode access through a compact permutation key that reconfigures a small subset of the model's parameters. We evaluate three private capabilities at the $180$M and $650$M scales, covering language acquisition, instruction following, and synthetic fact memorization. In each case, the keyed configuration cleanly acquires the private capability while the public configuration remains effectively unchanged, with a permutation key $560\times$ to $7{,}000\times$ smaller than a matched LoRA adapter. Crucially, neither fine-tuning the public configuration on related private data nor knowing up to $90\%$ of the key recovers the gated capability.

Taken together, these results position TLMs as an access-control primitive for open-weight models. A released checkpoint remains usable in its public configuration, while a compact structural key enables additional behavior for authorized users. Real deployments will require careful key management, larger-scale validation, stronger adaptive attacks, and evaluation on more realistic restricted capabilities, as discussed in Appendix~\ref{app:limitations}. Still, TLMs change the design space for open-weight release by showing that selective access can be built into a single, unified model, rather than enforced by withholding weights and mediating access through closed-source APIs.

\section{Limitations}
\label{app:limitations}

Our results show that a keyed parameter configuration can separate public and private behavior in controlled settings, but several important limitations remain.

\paragraph{Scale and capability realism.}
Our experiments are limited to 180M and 650M parameter models. This scale is sufficient to test whether permutation-keyed configurations can be trained at all, but it does not establish that the same design will behave identically in frontier-scale LLMs. Larger models may exhibit different interference patterns between public and keyed configurations, or expose new forms of leakage under white-box analysis. Similarly, we evaluate a fixed key size and a specific choice of swappable attention-head and MLP units. Future work could study whether the same tradeoff between separation, capacity, and efficiency holds across larger architectures, different key sizes, and longer contexts.

% \paragraph{Proxy capabilities and evaluation metrics.}
% The private tasks in this paper are controlled proxies, namely Spanish language modeling, instruction following, and synthetic fact memorization. These tasks cover distributional, behavioral, and pointwise forms of capability acquisition, but they are not the high-risk scientific, cyber, or proprietary-data settings that motivate TLMs. We also do not experiment on genuinely dangerous capabilities, so our results should not be read as evidence that TLMs are ready to gate real dual-use expertise. Validation loss, win rate, and exact-match recall provide clean diagnostics for whether a capability is isolated to the keyed configuration, but they do not measure the operational risks of deploying TLMs in sensitive domains. Future work should evaluate the framework on realistic restricted tasks using safe, redacted, or sandboxed benchmarks that better approximate real access-control demands.

\paragraph{Limited threat models.}
Our robustness experiments consider three attacks, namely fine-tuning the public configuration on partial private data, applying incomplete keys, and identifying tier parameters from weight statistics. These experiments serve as stress tests, but they do not cover the full space of adaptive white-box attacks. An adversary with substantial compute could attempt structured permutation search, activation-level analysis, or attacks that combine weight analysis with task knowledge. We also do not provide a cryptographic proof of security. The partial-key experiments show that recovery is not gradual in our setting, but they should not be interpreted as a formal guarantee.

\paragraph{Detectable fingerprints in the released weights.}
Private fine-tuning can leave statistical traces in the tier parameters. Our magnitude analysis shows that keyed units are partially distinguishable from non-keyed units after fine-tuning, especially in the MLP blocks. The simple attack we evaluate only recovers tier membership imperfectly and does not recover the permutation itself, but the existence of a fingerprint is still important. A stronger attack could use this signal to reduce the search space, combine it with other structural cues, or target future variants of the method. Reducing this fingerprint through better regularization, alternative key designs, or adversarially trained obfuscation is an important direction for making TLMs more robust.

\section*{Acknowledgments}

\noindent Arkil and Nicholas are partly supported by the Canada Graduate Scholarships (Doctoral) funded by the Natural Sciences and Engineering Research Council (NSERC) [funding reference no. 601601, 579783]. 
% Siva is supported by the Canada CIFAR AI Chairs program and the NSERC Discovery Grant program.
We thank the IVADO R10 AI Safety and Alignment regroupement for their generous support. 
We are grateful to Marius Mosbach and Ivan Titov for engaging in discussions during the early stages of this work. 
We would like to thank Shruti Joshi for providing helpful feedback on the technical writing and presentation aspects of this paper.
We thank our colleagues at Mila and McGill University for helpful discussions and for providing valuable feedback throughout this project.

\clearpage
\bibliography{references}
\bibliographystyle{plainnat}

\newpage
\appendix

% \section{Training Algorithms}

% \begin{algorithm}[h]
% \caption{Training a 2-tier Tiered Language Model}
% \label{alg:tlm_training}
% \begin{algorithmic}[1]
% \State Specify key $K$, yielding keyed configuration $\mathcal{C}_K$
% \State Set public configuration $\mathcal{C}_{\mathrm{pub}} = \mathrm{id}$
% \For{$t = 1$ to $T_{\mathrm{pre}}$}
%     \State Sample public batch $B_t \sim p_{\mathrm{pub}}$
%     \State Compute public loss $\ell_1 \leftarrow \mathrm{ForwardLoss}(\mathcal{M}_{\mathcal{C}_{\mathrm{pub}}(\theta)}; B_t)$
%     \State Compute keyed loss $\ell_2 \leftarrow \mathrm{ForwardLoss}(\mathcal{M}_{\mathcal{C}_{K}(\theta)}; B_t)$
%     \State Update $\theta_{\overline{S}}$ using $\lambda_1 \nabla_{\theta_{\overline{S}}}\ell_1 + \lambda_2 \nabla_{\theta_{\overline{S}}}\ell_2$
%     \State Update $\theta_{S}$ using $\nabla_{\theta_S}\ell_2$
% \EndFor
% \State Freeze $\theta_{\overline{S}}$ and cache public reference distributions from $\mathcal{M}^{\mathrm{pre}}_{\mathrm{pub}}$
% \For{$t = 1$ to $T_{\mathrm{ft}}$}
%     \State Sample private batch from $\mathcal{D}_{\mathrm{priv}}$ and reference batch from $p_{\mathrm{pub}}$
%     \State Compute $\mathcal{L}_{\mathrm{priv}}$ through $\mathcal{M}_{\mathcal{C}_K(\theta)}$
%     \State Compute public-behavior KL regularizer through $\mathcal{M}_{\mathcal{C}_{\mathrm{pub}}(\theta)}$
%     \State Update only $\theta_S$ using \cref{eq:ft_obj}
% \EndFor
% \end{algorithmic}
% \end{algorithm}

\section{Additional Details}\label{app:details}

\subsection{Multi-tier training details}
\label{app:multi_tier}

We formalize the multi-tier extension described in \cref{sec:multi_tier}. Let $K_1, \dots, K_N$ denote $N$ keys, each associated with a disjoint tier-parameter subset $S_i \subset \theta$, and let $\overline{S} = \theta \setminus \bigcup_{i=1}^{N} S_i$ denote the complementary parameters. Configurations are cumulative: $\mathcal{C}_i$ applies the permutations specified by $K_1, \dots, K_i$, so that $\mathcal{C}_i(\theta) = (K_i \circ \cdots \circ K_1)(\theta)$. Since each $K_j$ acts on a disjoint subset $S_j$, the composition order is immaterial.

\paragraph{Multi-tier pretraining.}
At each step, the public configuration $\mathcal{C}_0$ is always active and one keyed configuration $\mathcal{C}_i$ is selected by round-robin over $\{1, \dots, N\}$. When $\mathcal{C}_i$ is selected, tier parameters at or below tier $i$ receive gradients only from the keyed path, while all other parameters receive a mixture from both:
\begin{align}
\nabla_{\theta_{S_{\le i}}} \mathcal{L}_{\mathrm{pre}}
&\coloneqq
\mathbb{E}_{(x,y)\sim \mathcal{D}_{\mathrm{pub}}}
\Big[
\nabla_{\theta_{S_{\le i}}}\,\ell\big(p_{\mathcal{M}_{\mathcal{C}_{i}(\theta)}}(\cdot \mid x), y\big)
\Big],
\label{eq:multi_pretrain_s}
\\
\nabla_{\theta_{\overline{S} \cup S_{>i}}} \mathcal{L}_{\mathrm{pre}}
&\coloneqq
\mathbb{E}_{(x,y)\sim \mathcal{D}_{\mathrm{pub}}}
\Big[
\lambda_1 \nabla_{\theta_{\overline{S} \cup S_{>i}}}\,\ell\big(p_{\mathcal{M}_{\mathcal{C}_{0}(\theta)}}(\cdot \mid x), y\big)
+
\lambda_2 \nabla_{\theta_{\overline{S} \cup S_{>i}}}\,\ell\big(p_{\mathcal{M}_{\mathcal{C}_{i}(\theta)}}(\cdot \mid x), y\big)
\Big],
\label{eq:multi_pretrain_comp}
\end{align}
where $S_{\le i} = \bigcup_{j=1}^{i} S_j$ and $S_{>i} = \bigcup_{j=i+1}^{N} S_j$. The logic mirrors the two-tier case: parameters that the selected configuration permutes are shaped exclusively by that configuration, forcing the public path to learn around them.

\paragraph{Sequential multi-tier fine-tuning.}
Private fine-tuning proceeds sequentially: tier $1$ is fine-tuned first, then tier $2$ starting from the tier-$1$ checkpoint, and so on. When fine-tuning tier $i$ on $\mathcal{D}_i$, only $S_i$ is updated through $\mathcal{C}_i$. The objective consists of three terms:
\begin{multline}
\label{eq:multi_ft}
\mathcal{L}^{(i)}_{\mathrm{ft}}(\theta_{S_i})
=
\underbrace{
\mathbb{E}_{(x,y)\sim \mathcal{D}_{i}}
\big[
\ell\big(p_{\mathcal{M}_{\mathcal{C}_{i}(\theta)}}(\cdot \mid x), y\big)
\big]
}_{\text{private capability}}
+\;
\beta_{\mathrm{pub}} \underbrace{
\sum_{j=0}^{N}
\mathbb{E}_{x \sim \mathcal{D}_{\mathrm{pub}}}
\big[
\mathrm{KL}\big(
p_{\mathcal{M}_{\mathcal{C}_{j}(\widehat{\theta}_{\mathrm{pre}})}} \,\Vert\,
p_{\mathcal{M}_{\mathcal{C}_{j}(\theta)}}
\big)
\big]
}_{\text{public-behavior anchor}}
\\
+\;
\beta_{\mathrm{tier}} \underbrace{
\sum_{j=1}^{i-1}
\mathbb{E}_{x \sim \mathcal{D}_{j}}
\big[
\mathrm{KL}\big(
p_{\mathcal{M}_{\mathcal{C}_{j}(\widehat{\theta}^{(j)}_{\mathrm{ft}})}} \,\Vert\,
p_{\mathcal{M}_{\mathcal{C}_{j}(\theta)}}
\big)
\big]
}_{\text{earlier-tier preservation}},
\end{multline}
where $\widehat{\theta}^{(j)}_{\mathrm{ft}}$ denotes the checkpoint saved immediately after fine-tuning tier $j$. The first term trains the active tier on its private data. The second term generalizes the two-tier KL regularizer (\cref{eq:kl_reg}), anchoring \emph{all} configurations to their pretrained public-domain behavior. The third term is unique to the multi-tier setting: it prevents fine-tuning tier $i$ from degrading the private capabilities already acquired by earlier tiers, by anchoring each lower-tier configuration on its own private data against the reference distribution saved at the end of that tier's fine-tuning stage. \cref{fig:result_multi_tier} shows our full training curves throughout the stages of multi-tier fine-tuning.

\begin{figure}[t]
    \centering
    \includegraphics[width=\linewidth]{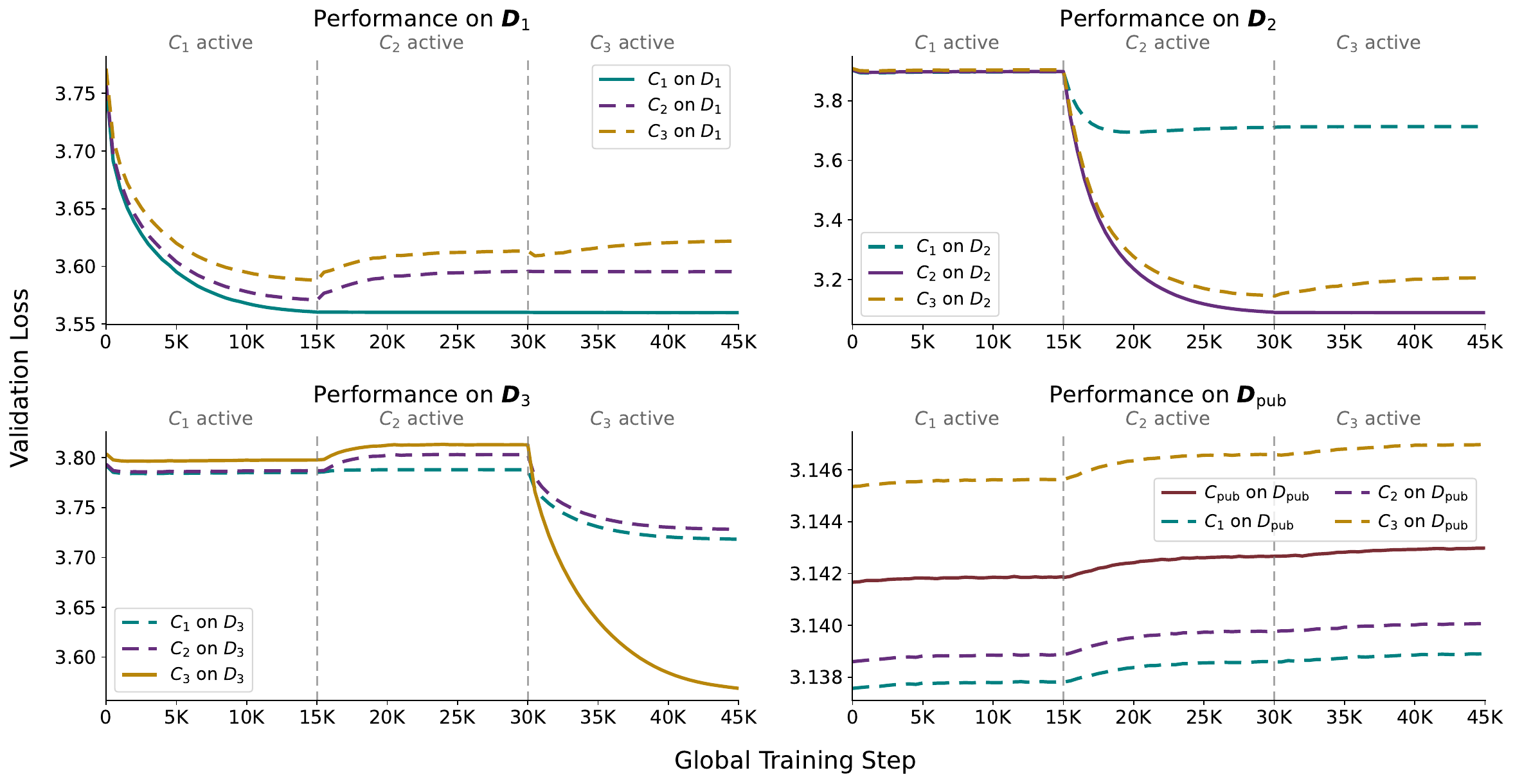}
    \caption{We start from the 180M cumulative multi-tier pretrained model with three 5\% keys. The private datasets are $ D_1=\mathrm{deu}, D_2=\mathrm{tur}, D_3=\mathrm{spa}. $  Each stage is trained on $2$B private tokens.}
    \label{fig:result_multi_tier_appendix}
\end{figure}

\subsection{Implementation Details}
\label{appx:training_hparams}

All TLMs are decoder-only GPT-Neo-style transformers trained in PyTorch \citep{paszke2019pytorch} with AdamW ($\beta_1=0.9$, $\beta_2=0.95$, weight decay $0.1$), a cosine learning-rate schedule decaying to a minimum value, and gradient clipping at norm $1.0$. We use bf16 mixed precision and standard PyTorch FSDP across 8 NVIDIA H100 80GB GPUs for every run. The token-to-parameter ratio at pretraining is $\approx 100$ for both scales, i.e., 18B tokens for TLM-180M and 65B tokens for TLM-650M.

\begin{figure}[t]
    \centering
    \includegraphics[width=0.8\linewidth]{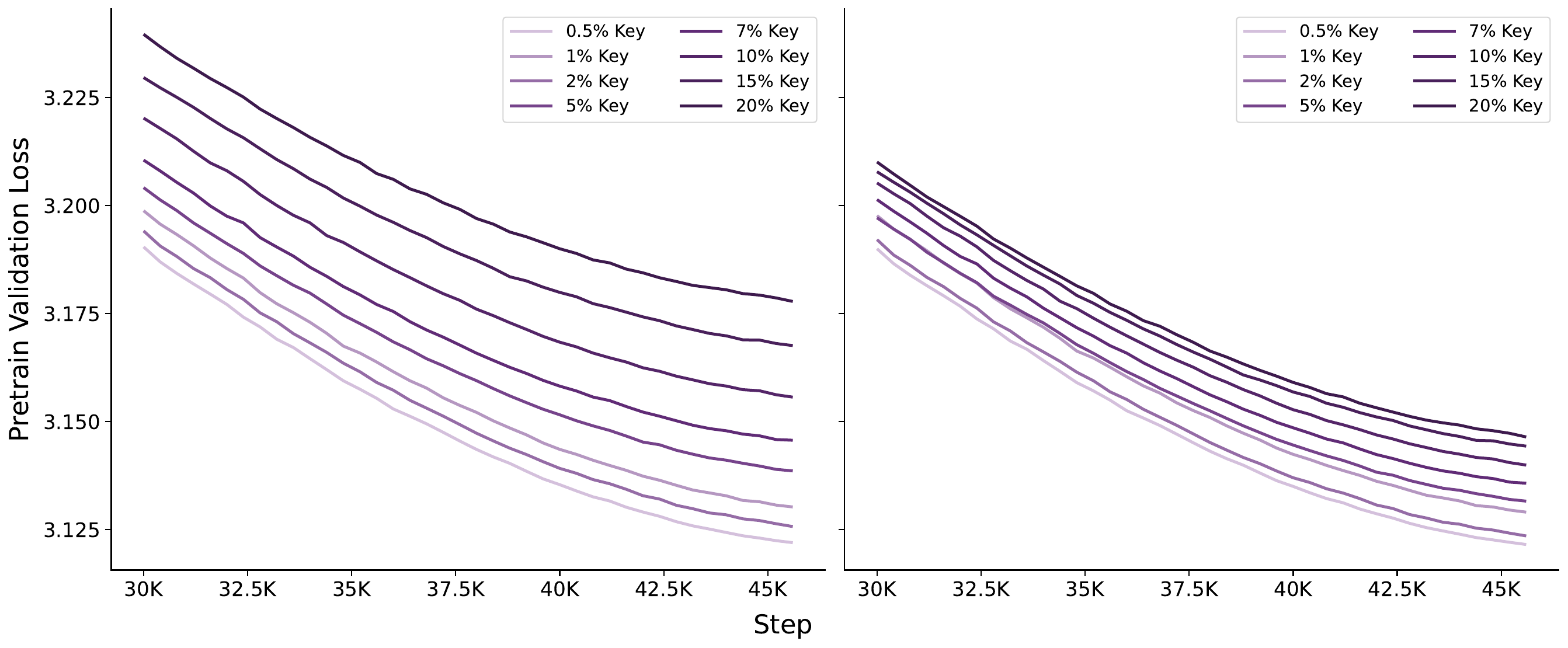}
    \caption{\textbf{Pretraining trajectories for the 180M TLM for all key-sizes.} \emph{Left}: public-configuration validation loss under $\mathcal{C}_{\mathrm{pub}}$. \emph{Right}: keyed-configuration validation loss under $\mathcal{C}_{K}$. Across key sizes, both configurations continue improving during pretraining, with larger keys showing slightly higher validation loss.}
    \label{fig:180m-pretrain-traj}
\end{figure}

\begin{figure}[t]
    \centering
    \includegraphics[width=0.8\linewidth]{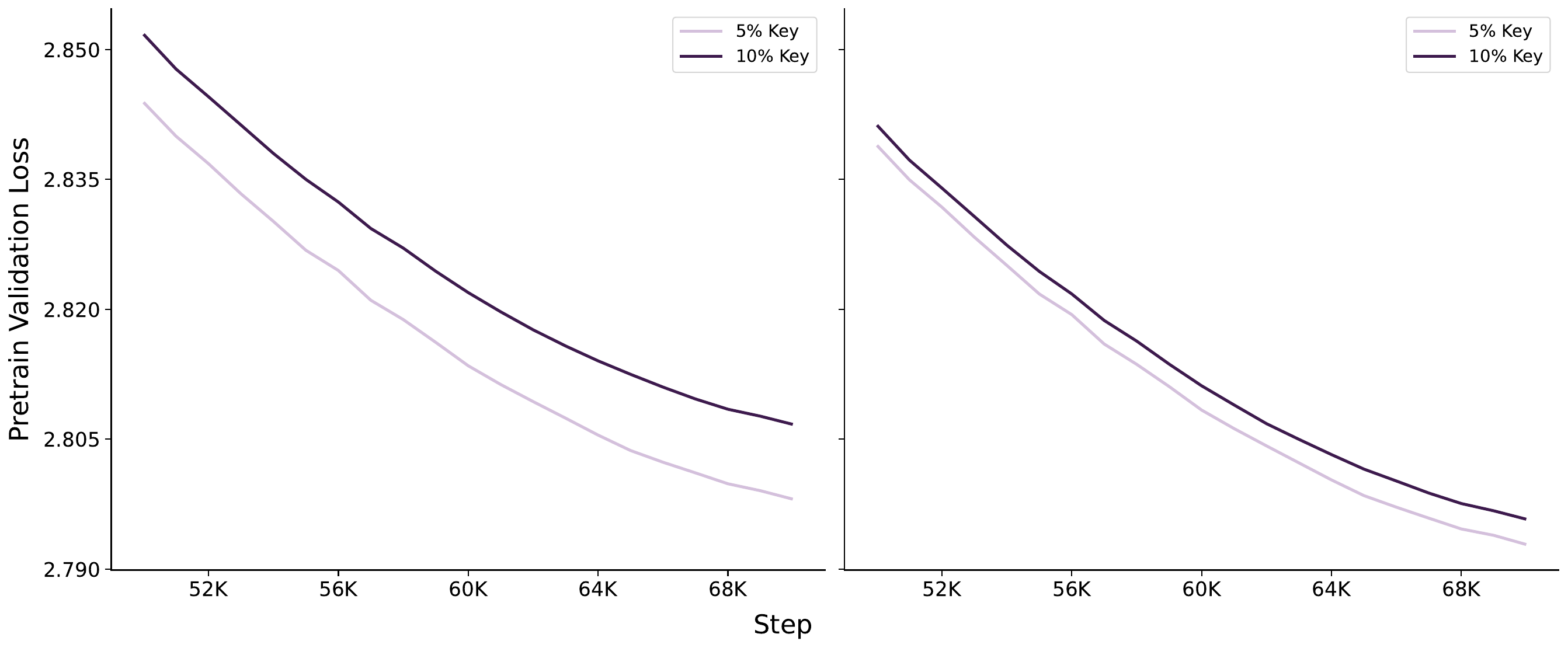}
    \caption{\textbf{Pretraining trajectories for the 650M TLM at two key-sizes.} \emph{Left}: public-configuration validation loss under $\mathcal{C}_{\mathrm{pub}}$. \emph{Right}: keyed-configuration validation loss under $\mathcal{C}_{K}$. Both key sizes continue improving throughout pretraining, with the 5\% key reaching slightly lower validation loss than the 10\% key on both configurations, consistent with the 180M trend.}
    \label{fig:650m-pretrain-traj}
\end{figure}

\begin{figure}[t]
    \centering
    \includegraphics[width=0.5\linewidth]{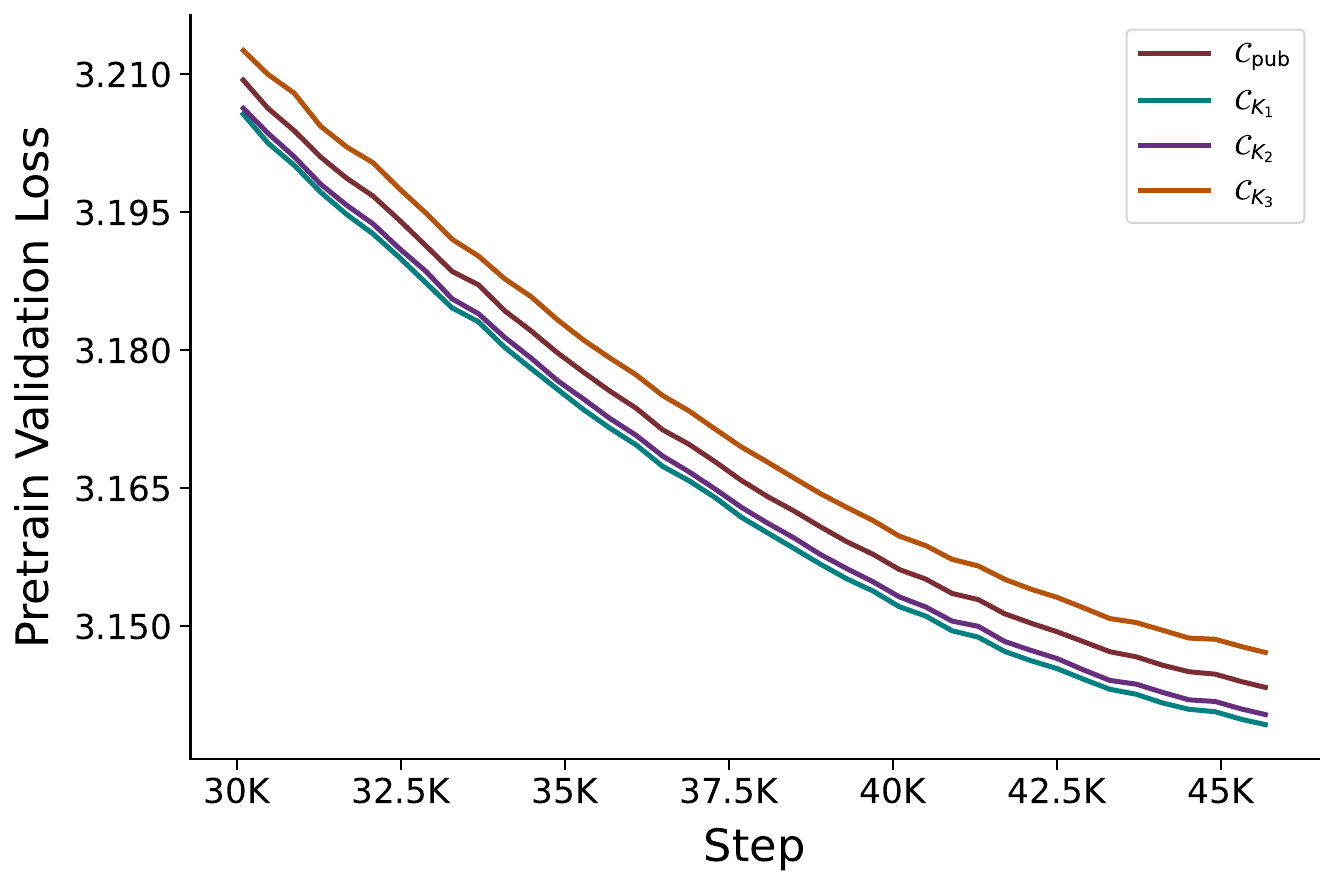}
    \caption{
    \textbf{Cumulative multi-tier pretraining.}
    A shared model is trained with one public configuration $\mathcal{C}_{\mathrm{pub}}$ and three nested keyed configurations: $\mathcal{C}_{K_1}$ applies key 1, $\mathcal{C}_{K_2}$ applies keys $1{+}2$, and $\mathcal{C}_{K_3}$ applies keys $1{+}2{+}3$. At each step, training uses $\mathcal{C}_{\mathrm{pub}}$ and one round-robin keyed configuration.
    }
    \label{fig:result_multi_tier_pretraining}
\end{figure}

\paragraph{Pretraining.} All pretraining runs use FineWeb, the asymmetric joint pretraining scheme of \cref{sec:method}, and the same keyed-update frequency $f=1$ unless otherwise noted. The pretraining loss mixing weights are $\lambda_1 = \lambda_2 = 0.5$ on the keyed parameters $\overline{S}$. \cref{tab:hparams_pretrain} reports the remaining details and both \cref{fig:180m-pretrain-traj} and \cref{fig:650m-pretrain-traj} show each model's pretraining trajectories under different key sizes. \cref{fig:result_multi_tier_pretraining} shows them for the cumulative pretraining.

\begin{figure}[t]
    \centering
    \begin{subfigure}[t]{0.32\linewidth}
        \centering
        \includegraphics[width=\linewidth]{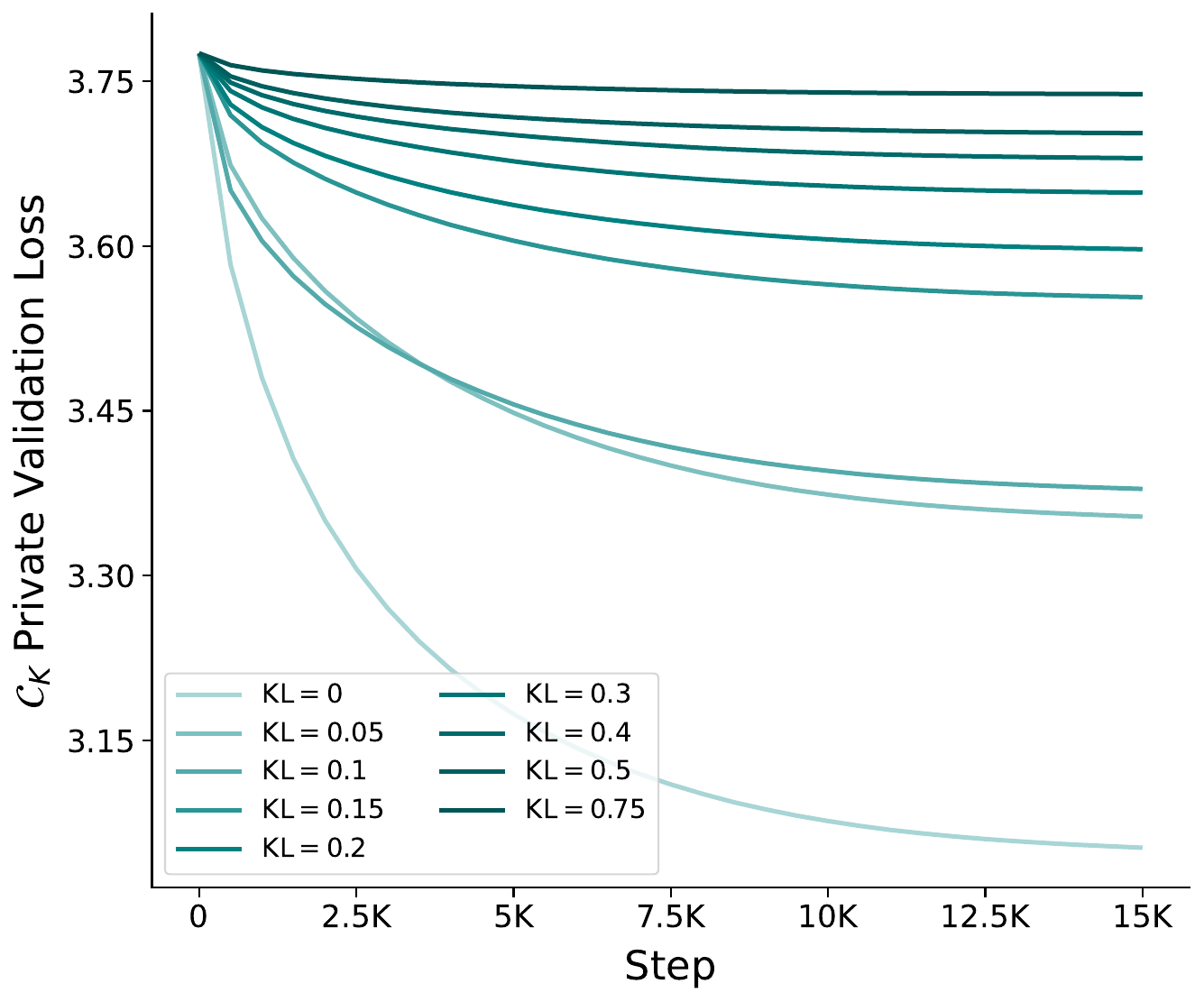}
        \caption{}
        \label{fig:kl_sweep_private}
    \end{subfigure}
    \hfill
    \begin{subfigure}[t]{0.32\linewidth}
        \centering
        \includegraphics[width=\linewidth]{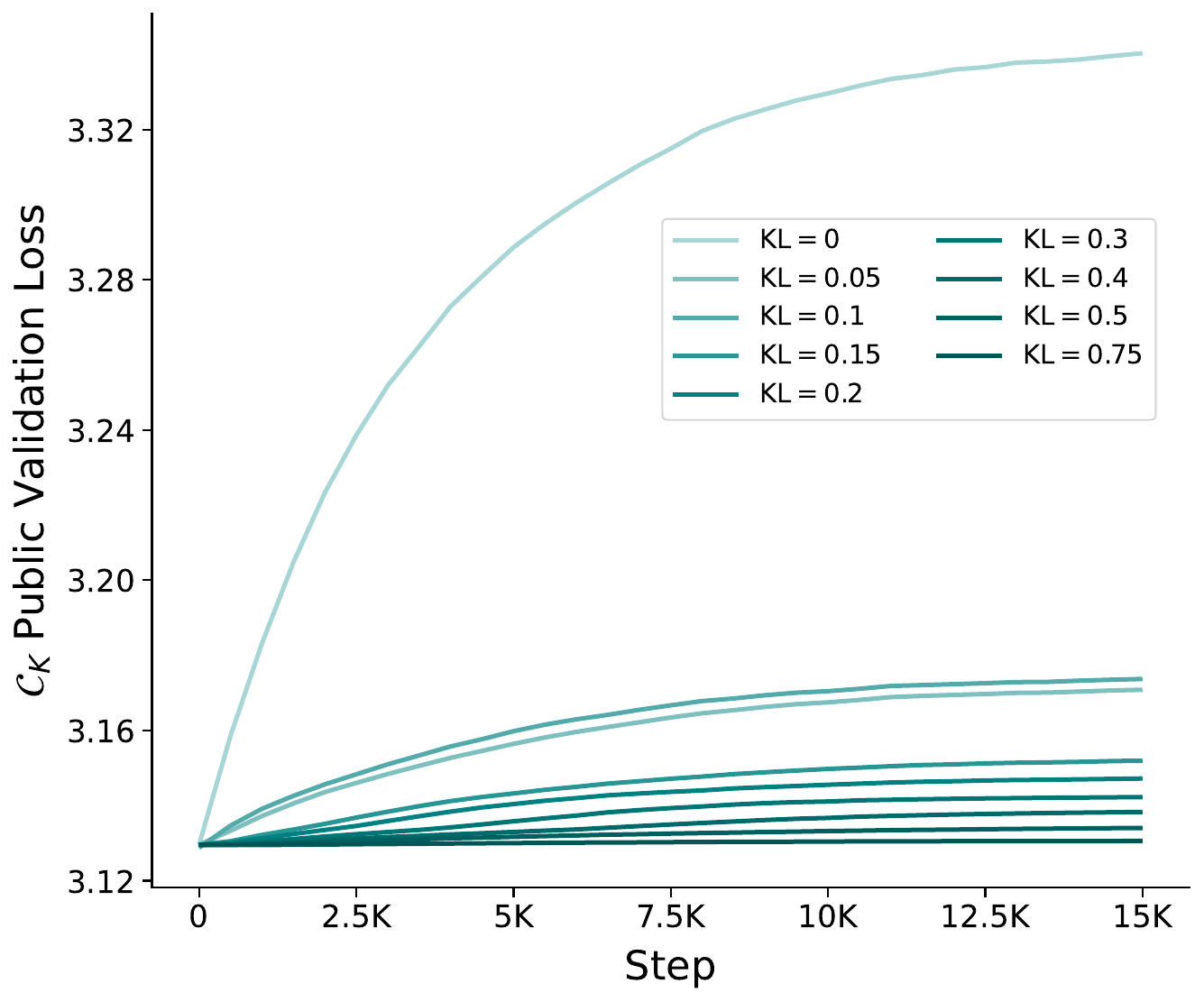}
        \caption{}
        \label{fig:kl_sweep_public}
    \end{subfigure}
    \hfill
    \begin{subfigure}[t]{0.32\linewidth}
        \centering
        \includegraphics[width=\linewidth]{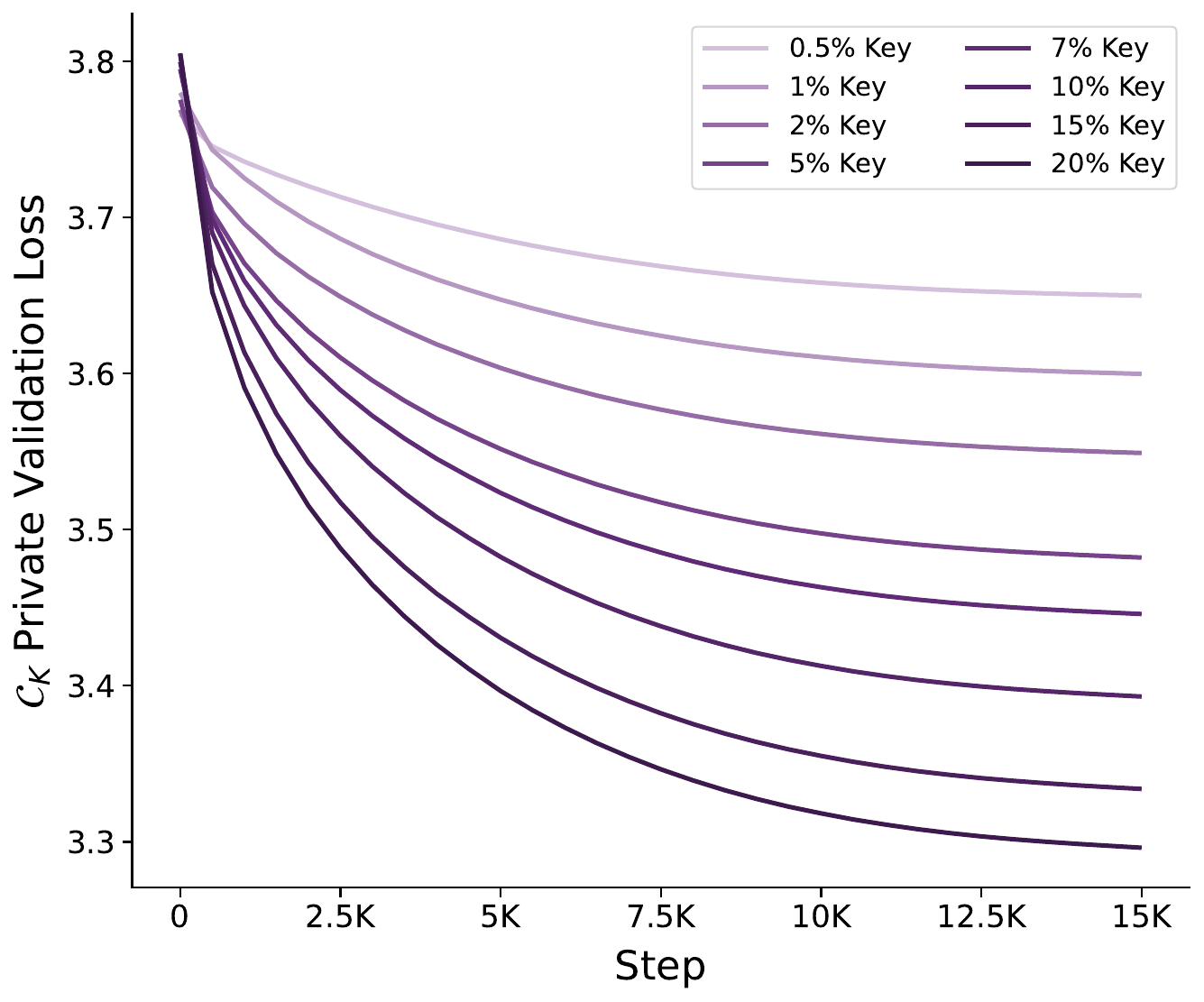}
        \caption{}
        \label{fig:key_size_sweep}
    \end{subfigure}
    \caption{\textbf{KL and key-size sweeps during private fine-tuning of a 180M model on 2B tokens of FineWeb2 Spanish.} Weaker KL regularization lets $\mathcal{C}_{K}$ adapt more strongly to the private distribution (\subref{fig:kl_sweep_private}) at the cost of greater drift from previously learned public behavior (\subref{fig:kl_sweep_public}); darker curves correspond to larger KL weights. Larger key fractions yield lower private validation loss (\subref{fig:key_size_sweep}); darker curves correspond to larger key sizes.}
    \label{fig:result_kl_and_key_sweep}
\end{figure}

\paragraph{Private fine-tuning.} All fine-tuning runs share the objective in \cref{eq:ft_obj}: the keyed-configuration cross-entropy loss on private data plus a public-behavior KL regularizer with $\beta_{\mathrm{KL}}=0.1$ (we include a $\beta$ sweep in \cref{fig:result_kl_and_key_sweep}). We additionally apply a small $\ell_2$ penalty ($\lambda_{S}=0.01$) on the keyed parameters to discourage runaway updates. All share a 5\% key unless otherwise specified (we also include a fine-tuning sweep for various key sizes in \cref{fig:key_size_sweep}). Only the tier parameters $S$ are trainable; the complementary set $\overline{S}$ is frozen at its pretrained values. We use the same AdamW / cosine schedule / bf16 / 8-GPU FSDP setup as in pretraining.  \cref{tab:hparams_finetune} collects the per-run details. The "token budget" column gives the target private-token count when applicable; for fixed-epoch runs (synbios, Alpaca), we instead report the nominal step count.

\begin{table}[h]
\centering
\caption{Pretraining hyperparameters for TLM-180M, TLM-650M, and the
three-tier cumulative TLM-180M.}
\label{tab:hparams_pretrain}
\setlength{\tabcolsep}{4pt}
\begin{tabular}{@{}lccc@{}}
\toprule
 & 180M & 650M & 180M 3-tier \\
\midrule
GPUs & 8 & 8 & 8 \\
Per-GPU batch & 24 & 14 & 24 \\
Grad. accumulation & 1 & 4 & 1 \\
Global batch & 192 & 448 & 192 \\
Sequence length & 2048 & 2048 & 2048 \\
Tokens/step & 393{,}216 & 917{,}504 & 393{,}216 \\
Peak LR & $4.2{\times}10^{-4}$ & $2.8{\times}10^{-4}$ & $4.2{\times}10^{-4}$ \\
Min LR & $4.2{\times}10^{-5}$ & $2.8{\times}10^{-5}$ & $4.2{\times}10^{-5}$ \\
Warmup steps & 1{,}000 & 1{,}000 & 1{,}000 \\
Total steps & 45{,}776 & 70{,}844 & 45{,}776 \\
Total tokens & 18B & 65B & 18B \\
Runtime & $\sim$12H & $\sim$3.5D & $\sim$16H \\
Eval interval & 400 & 1{,}000 & 400 \\
Number of keys & 1 & 1 & 3 \\
Per-key size & 5\% & 5\% & 5\% each \\
Tier sampling & --- & --- & round-robin \\
\bottomrule
\end{tabular}
\end{table}

\begin{table}[h]
\centering
\small
\caption{Private fine-tuning hyperparameters across all downstream
settings.}
\label{tab:hparams_finetune}
\setlength{\tabcolsep}{4pt}
\begin{tabular}{lccccc}
\toprule
 & 180M (lang.) & 180M Syn-bios & 650M (lang.) & 650M Alpaca & 180M Multi \\
\midrule
Dataset & spa &  synth.\ bios & spa/tur/por  & Alpaca & deu, tur, spa \\
GPUs & 8 & 8 & 8 & 8 & 8 \\
Per-GPU batch size & 8 & 8 & 4 & 4 & 8 \\
Sequence length & 2048 & 2048 & 2048 & 1024 & 2048 \\
Peak learning rate & $1\!\times\!10^{-5}$ & $3\!\times\!10^{-5}$ & $1\!\times\!10^{-5}$ & $1\!\times\!10^{-5}$ & $1\!\times\!10^{-5}$ \\
Min learning rate & $1\!\times\!10^{-6}$ & $1\!\times\!10^{-6}$ & $1\!\times\!10^{-6}$ & $1\!\times\!10^{-6}$ & $1\!\times\!10^{-6}$ \\
Warmup steps & 100 & 100 & 100 & 100 & 100 (per stage) \\
Total steps & 15{,}259 & 4{,}050 & 61{,}035 & 4{,}773 & 45{,}777 \\
Token budget & 2B & $\sim$7M & 4B & $\sim$157M  & 6B \\
Runtime & $\sim$2.5H & $\sim$0.5H & $\sim$10H & $\sim$1.5H  & $\sim$27.5H \\
$\beta_{\mathrm{KL}}$ (public) & 0.1 & 0.1 & 0.1 & 0.1 & 0.1 \\
$\beta_{\mathrm{anchor}}$ (lower tiers) & --- & --- & --- & --- & 0.1 \\
$\lambda_{S}$ & 0.01 & 0.01 & 0.01 & 0.01 & 0 \\
Eval interval (steps) & 500 & 50 & 500 & 500 & 500 \\
\bottomrule
\end{tabular}
\end{table}

% \section{Broader impact}
% \label{app:broader-impact}

% TLMs are motivated by the potential social benefits of open-weight models: broader scientific access, easier self-hosting in privacy-sensitive settings, and reduced reliance on closed API mediation for authorized use. If effective, a tiered model could make it easier to share a common public checkpoint while selectively enabling additional capabilities for trusted users, reducing the need to choose between fully withholding weights and releasing all capabilities equally.

% At the same time, TLMs introduce security and governance risks. If deployed before their robustness is well understood, they could create a false sense of protection around genuinely dangerous capabilities; if the mechanism fails, private behavior could leak or become recoverable through stronger adaptive attacks than those studied here. The mechanism could also be misused to hide capabilities in released models, and leaked keys cannot be centrally revoked once both weights and keys are in the wild. We recommend future deployments use domain-specific risk evaluation, adaptive red-teaming, careful key management, and additional release controls.

\subsection{Constructing the synthetic biography dataset}
\label{appx:synbios}

We construct a synthetic biography dataset containing $400$ fictitious people, each defined by four unique attributes: age, profession, hobby, and salary. Names are drawn from a curated pool of $200$ male and $200$ female first names; professions are drawn from a pool of $400+$ distinct occupations annotated with the correct indefinite article (\emph{a} or \emph{an}); hobbies are drawn from a pool of $400+$ short single-word activities (e.g.\ \texttt{swimming}, \texttt{calligraphy}); and salaries are sampled as integer dollar amounts in $[\$25{,}000, \$425{,}000]$. Professions, hobbies, and salaries are sampled without replacement, so that no two people share any of these three attributes; ages are sampled independently in $[22, 85]$. Salaries are deliberately not constrained to round-number multiples, so that exact-match recall cannot be achieved by predicting common $\$XX{,}000$ tokens.

Each biography is generated from four short templates encoding age, profession, hobby, and salary. The first sentence uses the person's name and later sentences use the gendered pronoun (\texttt{He}/\texttt{She}). For each person, we include all $4! = 24$ permutations of the four statements to remove ordering cues, producing $9{,}600$ biographies in total. A typical example is:

\emph{"Alice works as a Doctor. She is 42 years old. She enjoys swimming. She earns \$83{,}472."}

We fine-tune the $180$M TLM on this dataset for approximately $27$ epochs. At evaluation, we prompt the model with the first three statements of the biography and ask it to predict the target attribute value in the fourth statement, decoding the continuation greedily. We report two metrics, averaged across all $24$ permutations per person: \emph{exact match}, defined as $1$ if the decoded continuation matches the target attribute value token-for-token and $0$ otherwise; and \emph{partial match}, defined as the fraction of target tokens that the greedy decode predicts correctly at the matching positions. Exact match is the strict criterion used in the memorization experiment in \cref{fig:result_cost_memorization} (left).

\section{Clarifications}

\noindent \textbf{(1)} \textit{Why not maintain separate model variants or LoRA adapters for different capability tiers?}

Maintaining capability tiers as distinct model variants introduces a systems burden. Modern serving stacks are most efficient when requests share weights, memory pools, and batching structure, whereas heterogeneous checkpoints or adapter-specialized variants create duplication, fragmented GPU allocation, and weaker batching. LoRA reduces the cost of training and storing such variants, but not the cost of serving them at scale: systems must still load, schedule, and batch adapter-specific computations with different ranks, request lengths, and memory footprints. This challenge has motivated specialized LoRA-serving systems, alongside broader multi-model serving systems for multiplexing distinct checkpoints under bursty demand \citep{sheng2023s,chen2024punica,10.1145/3731569.3764815,duan2024muxserve}. The hope is that future work can leverage TLMs to sidestep this deployment layer by representing access tiers as configurations of a single checkpoint, rather than separate models or adapter modules. Additionally, as shown in \cref{tab:key_storage}, our approach is much more efficient in terms of storage, and potentially, transmission, compared to LoRA. Finally, hiding adapter weights corresponding to certain private capabilities goes against the spirit of open science, and prevents community research on topics such as interpretability.

\noindent \textbf{(2)} \textit{Why do we use validation loss as a central metric, and does lower loss imply useful private capability?}

Validation loss is not intended to be a universal measure of model usefulness. Here, we use it for the narrower goal of testing whether a configuration has adapted to a target distribution while preserving performance on the public distribution. In the Spanish data experiment (\cref{fig:result_behavioral_separation} (left)), the private capability is language modeling on Spanish text, making private-domain validation loss a direct measure of how well each configuration models that domain. Public-domain validation loss measures whether private fine-tuning degrades the default public model. Since these comparisons use the same architecture, tokenizer, data distribution, and evaluation protocol, loss provides a controlled diagnostic for separation and preservation.

\noindent \textbf{(3)} \textit{How should we interpret seemingly modest differences in validation loss?}

Validation loss is a dense, token-averaged quantity, and its numerical scale should be interpreted comparatively. In language-model scaling work, held-out cross-entropy is the standard signal used to compare models across size, data, and compute budgets, and smooth changes in loss are predictive of meaningful differences in model quality~\citep{hoffmann2022training}. This compression is visible even in our own pretraining runs: the $180$M TLM reaches a public-domain validation loss of $3.13161$ under $\mathcal{C}_{K}$, while the $650$M TLM reaches $2.79283$. The difference is only $0.33878$ nats, yet the larger model is substantially more capable.

For this reason, we interpret our loss curves comparatively and by configuration. The same checkpoint improves on the private distribution under $\mathcal{C}_{K}$ while remaining nearly unchanged under $\mathcal{C}_{\mathrm{pub}}$, and public-domain loss remains stable during private fine-tuning.

\noindent \textbf{(4)} \textit{What does evaluation at this model scale establish?}

Although our models are smaller than frontier LLMs, this scale is appropriate for isolating the mechanism. It allows us to train multiple tiered configurations, evaluate cross-tier interference, and run controlled ablations that would be prohibitively expensive at frontier scale. Our goal is to show that keyed parameter reconfiguration can induce distinct functional behavior in a single checkpoint. Evaluation on larger models remains an important direction for future work.

\section{Additional Results and Discussion}\label{app:results}

\subsection{Permuting the weights of a trained model destroys its capabilities}
\label{app:permutation_destroys}

A natural question is whether tiered pretraining is necessary at all. One might hope to take an off-the-shelf pretrained model and apply a permutation key post hoc. We show that this fails. Permuting even a small fraction of a trained transformer's parameters severely degrades its capabilities, since the learned computation depends on precise alignment between parameter positions across layers. Tiered pretraining is therefore necessary to achieve our goal.

\begin{figure}[t]
  \centering

  \begin{minipage}[t]{0.48\linewidth}
    \centering
    \includegraphics[width=\linewidth]{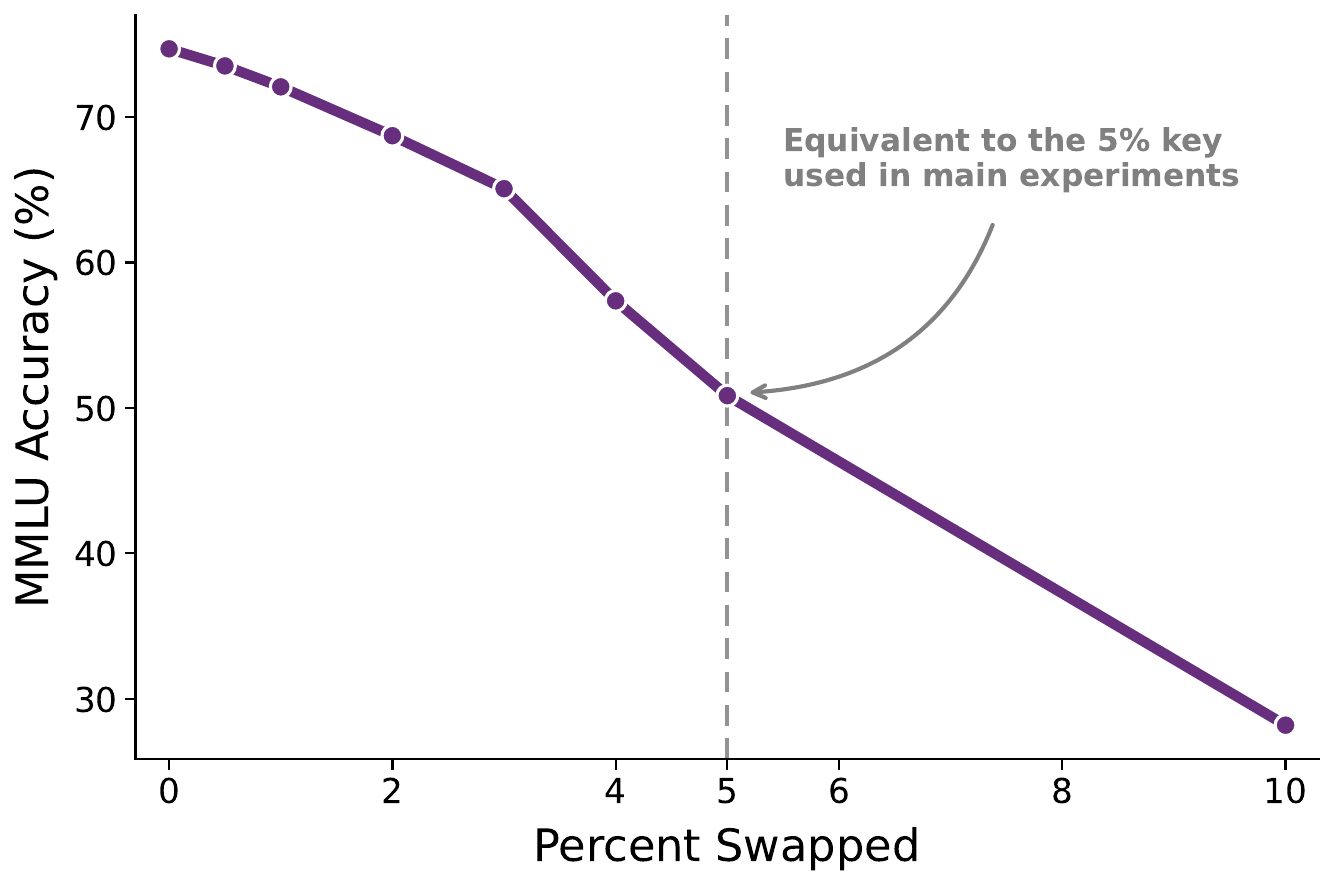}
  \end{minipage}
  \hfill
  \begin{minipage}[t]{0.48\linewidth}
    \centering
    \includegraphics[width=\linewidth]{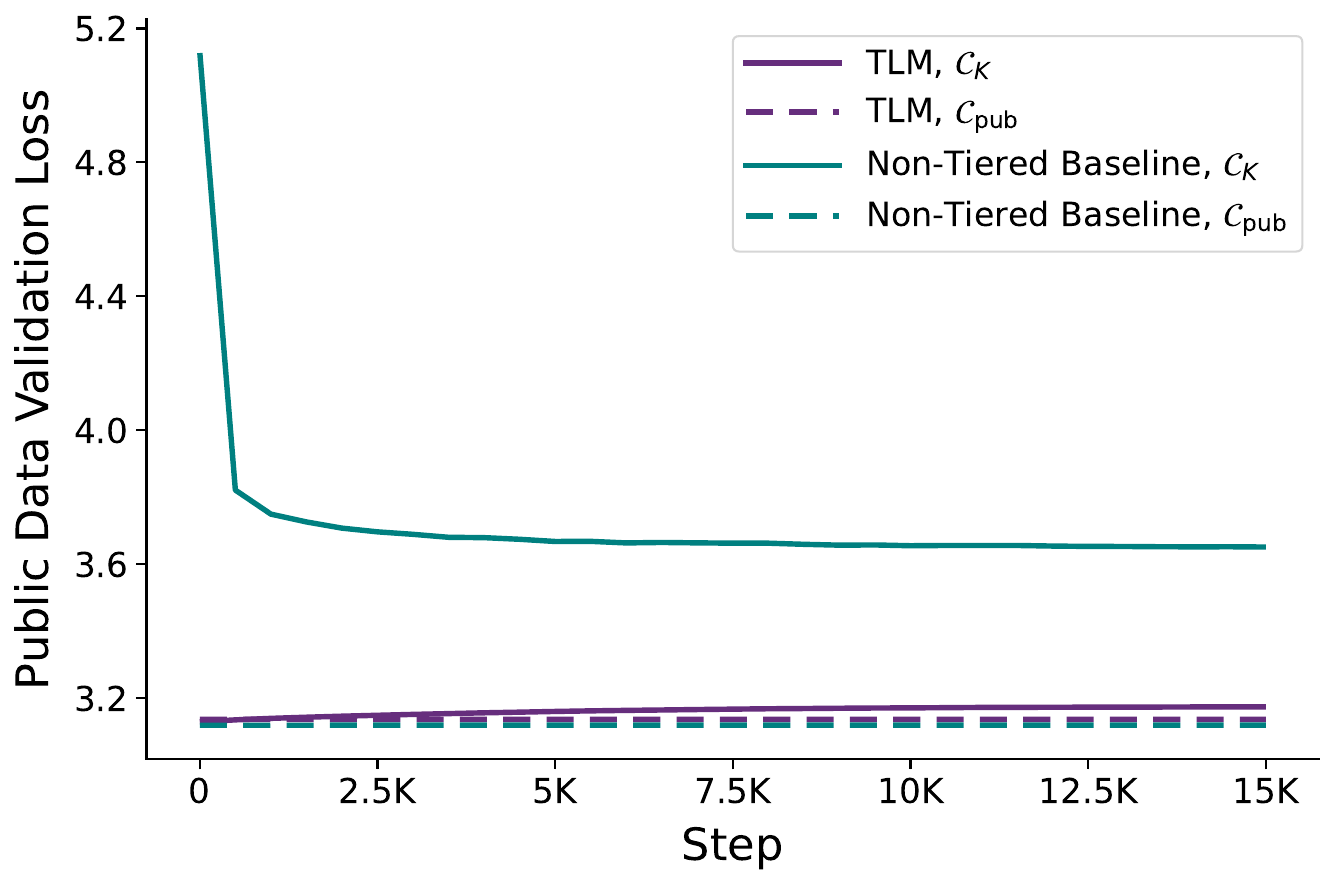}
  \end{minipage}

  \caption{
    \emph{Left:} \textbf{Permuting weights of a pretrained model destroys capabilities.} We apply random parameter permutations to Qwen-3-8B, allocating 25\% of the swap budget to attention heads and 75\% to MLP columns, and evaluate MMLU accuracy.
    \emph{Right:} \textbf{Tiered pretraining is necessary for the keyed configuration to function.}
    Public-domain validation loss during private fine-tuning on $2$B Spanish tokens, comparing a TLM and a non-tiered baseline under $\mathcal{C}_{\mathrm{pub}}$ (dashed) and $\mathcal{C}_K$ (solid). In the non-tiered baseline, $\mathcal{C}_K$ starts with very high loss and recovers only partially, whereas both TLM configurations remain stable.
  }
  \label{fig:qwen-and-baseline}
\end{figure}

\paragraph{Permuting a pretrained model.}
We take Qwen-3-8B \citep{yang2025qwen3technicalreport} and apply random weight permutations of increasing size, using the same swap structure as in our TLM experiments (25\% of the swap budget allocated to attention heads, 75\% to MLP columns). \Cref{fig:qwen-and-baseline} (left) reports MMLU accuracy as a function of the fraction of parameters involved in the permutation. Accuracy drops sharply as the permutation size increases. At a 5\% swap fraction (matching the key size used throughout our main experiments), MMLU falls from 74.7\% to 50.8\%. This is a nearly 24-point reduction from a perturbation that affects only a small fraction of the model, keeping in mind that 10\% accuracy is close to chance. These results show that post-hoc permutation is not a viable access-control mechanism for standard pretrained transformers. Without tiered pretraining, the key acts as a destructive perturbation rather than an alternate functional configuration.

\paragraph{Tiered pretraining prevents the degradation of public capabilities under the key.}
The previous experiment uses a model that was never trained to accommodate permutations. We next ask whether tiered pretraining resolves this fragility for the specific permutation it was trained with. We take two 180M-parameter models pretrained on the same data: a standard (non-tiered) baseline and a TLM. We then fine-tune both on 2B tokens of Spanish data, \emph{applying the keyed permutation $\mathcal{C}_K$ to both during fine-tuning}. \Cref{fig:qwen-and-baseline} (right) shows the public-domain validation loss under both configurations throughout fine-tuning. For the TLM, $\mathcal{C}_K$ and $\mathcal{C}_{\mathrm{pub}}$ both maintain stable, low public-domain loss, confirming that tiered pretraining has taught the model to function well under the keyed permutation. For the non-tiered baseline, applying $\mathcal{C}_K$ initially produces validation loss above $5.0$, indicating that the permutation has severely disrupted the learned computation. The loss recovers partially during fine-tuning as the keyed parameters are updated, but the baseline under $\mathcal{C}_K$ never reaches the public-domain quality of either TLM configuration, maintaining a significant gap. In contrast, the baseline under $\mathcal{C}_{\mathrm{pub}}$ remains unaffected, as expected, since it uses the original parameter arrangement.

Together, these results establish that the asymmetric joint pretraining stage of our method is essential. Without it, the keyed configuration would start from a broken model state, and fine-tuning alone \emph{cannot} fully recover the lost structure.

\subsection{Comparison against a non-tiered baseline}

% \begin{figure}[t]
%     \centering
%     \includegraphics[width=0.6\linewidth]{plots/180m_memo_synthetic.pdf}
%     \caption{\textbf{Memorization of synthetic facts.} Exact-match accuracy on a held-out set of synthetic biographies during private finetuning of TLM-180M. The keyed configuration $\mathcal{C}_K$ reaches perfect recall of all $400$ facts; the public configuration $\mathcal{C}_{\mathrm{pub}}$ remains at zero throughout, confirming that the tier parameters can store precise factual knowledge with no leakage into the public model.}
%     \label{fig:result_memorization}
% \end{figure}

\begin{figure}[t]
  \centering

  \begin{minipage}[t]{0.48\linewidth}
    \centering
    \includegraphics[width=\linewidth]{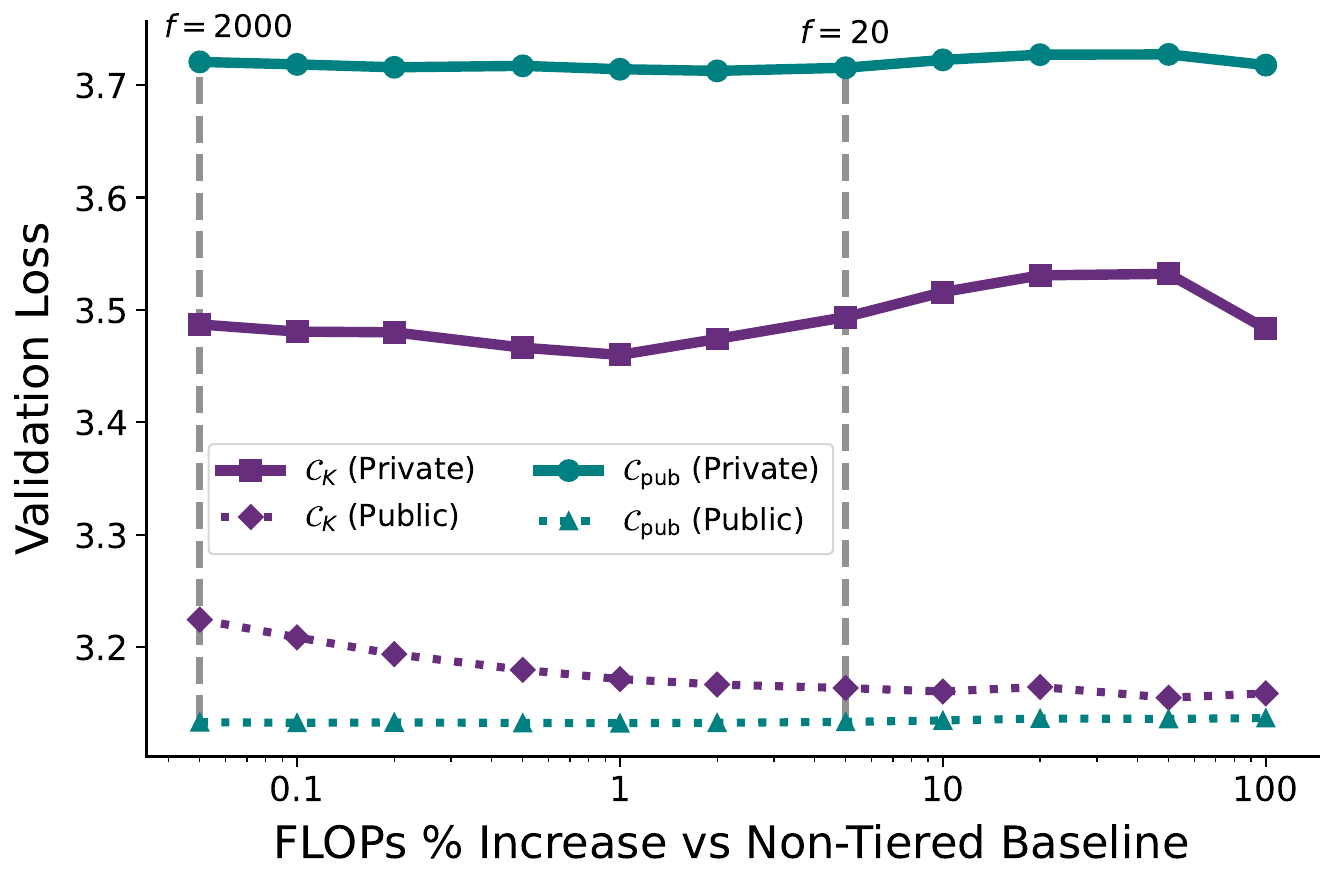}
  \end{minipage}
  \hfill
  \begin{minipage}[t]{0.48\linewidth}
    \centering
    \includegraphics[width=\linewidth]{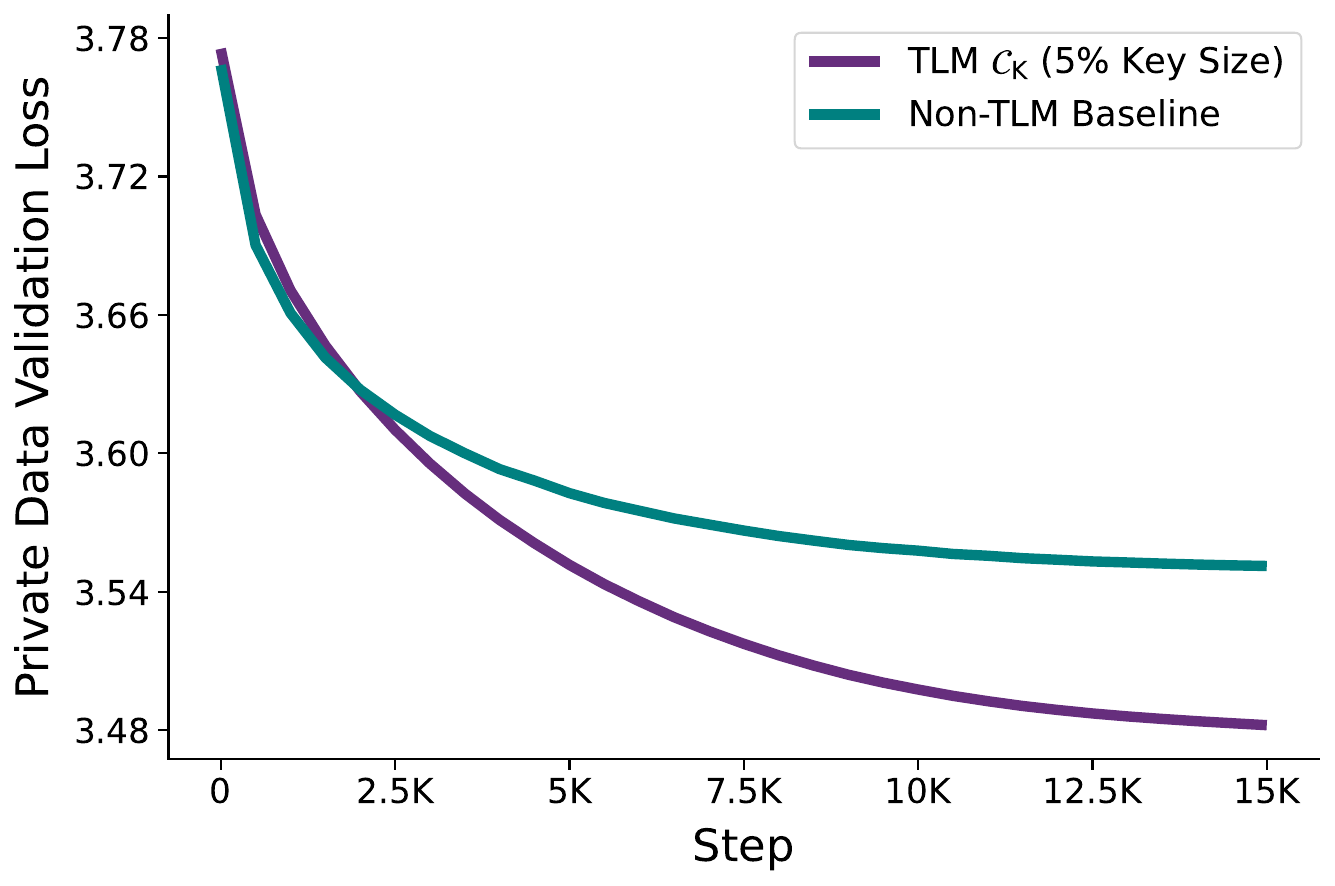}
  \end{minipage}

  \caption{
    \emph{Left:} \textbf{Effect of keyed-update frequency on fine-tuning.}
    Each $180$M TLM from \cref{fig:result_cost_memorization} (right) is fine-tuned on $2$B tokens of Spanish FineWeb2. $\mathcal{C}_K$'s private-domain loss stays nearly flat across values of $f$, while $\mathcal{C}_{\mathrm{pub}}$ remains high, indicating preserved behavioral separation.
    \emph{Right:} \textbf{Private-domain performance: TLM vs.\ Non-TLM Baseline.} Private-domain validation loss during fine-tuning on $2$B tokens of Spanish FineWeb2 data, comparing the keyed configuration of TLM-180M against a non-tiered baseline of the same architecture fine-tuned on the same data.
  }
  \label{fig:eff-and-baseline}
\end{figure}

% \begin{figure}[t]
%     \centering
%     \includegraphics[width=0.5\linewidth]{plots/tlm_vs_baseline.pdf}
%     \caption{\textbf{Tiered pretraining is necessary for the keyed configuration to function.} Public-domain validation loss during private finetuning on 2B tokens of Spanish data, comparing a TLM and a non-tiered baseline of the same 180M architecture. Both models are evaluated under $\mathcal{C}_{\mathrm{pub}}$ (dashed) and $\mathcal{C}_K$ (solid). For the non-tiered baseline, $\mathcal{C}_K$ starts at a validation loss above 5.0---the permutation has effectively destroyed the learned computation---and recovers only partially during finetuning. For the TLM, both configurations maintain stable low loss throughout, confirming that tiered pretraining produces a model whose internal representations are robust to the trained permutation.}
%     \label{fig:result_tlm_vs_baseline_permuted}
% \end{figure}

A natural question is how much tiered pretraining costs relative to standard training. To quantify this, we train a non-tiered language model of the same architecture under identical conditions and compare it against the $180$M TLM along two axes: public-domain quality after pretraining, and private-domain performance after fine-tuning.

\paragraph{Public-domain quality after pretraining.}
As discussed in \Cref{sec:analysis}, \Cref{fig:pretrain_comp} compares the public-domain validation loss of the TLM against the non-tiered baseline over the final portion of pretraining. The two curves are nearly parallel, with the TLM consistently trailing by a small horizontal offset: the TLM requires roughly $6\%$ more training steps to reach any given loss value that the baseline has already achieved. Given that tiered pretraining simultaneously prepares the keyed configuration for downstream private fine-tuning, we consider this a modest overhead.

\paragraph{Private-domain performance after fine-tuning.}
We next ask whether the keyed configuration can match a conventionally trained model on private data. We fine-tune both models on $2$B tokens of Spanish FineWeb2 data: the TLM is fine-tuned through $\mathcal{C}_K$ as described in \cref{sec:method}, and the non-tiered baseline is fine-tuned by updating the same parameter subset without any permutation. As shown in \cref{fig:eff-and-baseline} (left), the keyed TLM converges to a similar final private-domain loss as the baseline. This means that the TLM training approach does not limit the model's capacity to acquire private knowledge.

\subsection{Identifying tier parameters from weight magnitudes}
\label{app:weight_analysis}

\begin{figure}[t]
  \centering

  \begin{subfigure}{\linewidth}
    \centering
    \includegraphics[width=\linewidth]{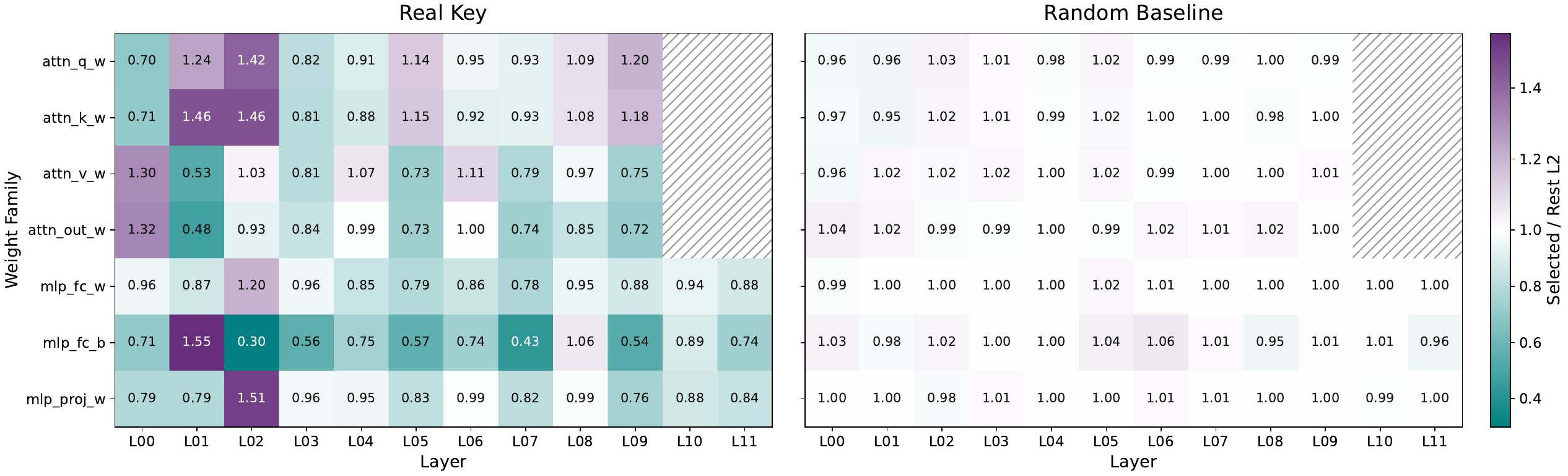}
    \caption{Spanish FineWeb2 fine-tuning.}
    \label{fig:spanish-heatmap}
  \end{subfigure}

  \begin{subfigure}{\linewidth}
    \centering
    \includegraphics[width=\linewidth]{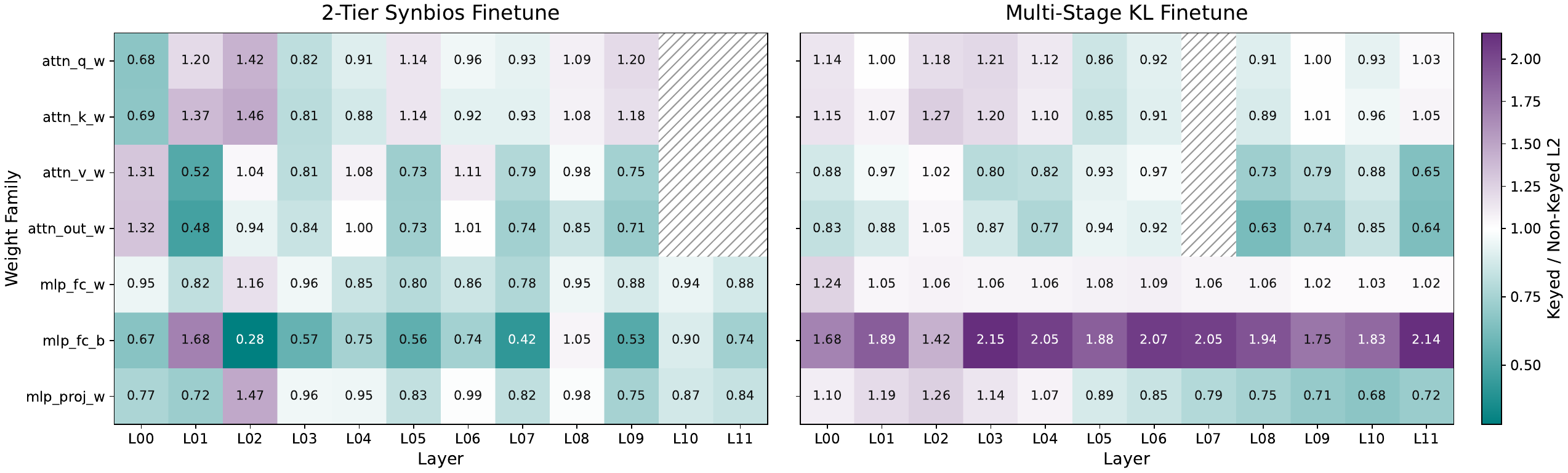}
    \caption{Synbios and multilingual multi-stage fine-tuning.}
    \label{fig:synbios-vs-multi}
  \end{subfigure}

  \caption{
    Weight-magnitude signatures of tier parameters after private fine-tuning.
    Each cell shows the ratio between the mean $L_2$ norm of key-selected units
    and size-matched unselected units for a given module family and layer.
    Values near $1.0$ indicate that the two groups are indistinguishable by magnitude.
    Random size-matched subsets remain close to $1.0$, showing that the observed
    structure is tied to tiered training rather than sampling noise.
  }
  \label{fig:magnitude-signatures}
\end{figure}

The tier parameters $S$ are updated differently compared to the rest of the weights $\overline{S}$. This may leave a statistical fingerprint in the released weights. We investigate whether an adversary can exploit this to identify which parameters belong to $S$.

\paragraph{Visualizing $\mathbf{S}$ vs. $\mathbf{\overline{S}}$ magnitudes}
We examine the $180$M TLM after private fine-tuning on Spanish data. For each module family (attention $q/k/v/o$ projections, MLP up/down projections, and biases) and each layer, we compute the ratio of the mean $L_2$ norm of key-selected units (attention heads or MLP dimension blocks, depending on the module) to that of unselected units. A ratio near $1.0$ means the two groups are indistinguishable by magnitude; deviations indicate a detectable signature. \Cref{fig:spanish-heatmap} (left) shows this ratio for the true key on the Spanish dataset, while \cref{fig:spanish-heatmap} (right) shows the same computation with randomly chosen, size-matched subsets as a control. The random control stays close to $1.0$ everywhere, confirming that the structure visible in the left panel is not a sampling artifact but a genuine consequence of our training paradigm. We repeat the same analysis for the synthetic-biography and multilingual multi-tier settings in \Cref{fig:synbios-vs-multi}. The Spanish and synthetic-biography fine-tuned models reveal a consistent magnitude fingerprint in which a significant portion of the keyed units, especially in the MLP blocks, tend to have smaller norms than non-keyed units. However, cumulative multi-tier fine-tuning partially alleviates this effect. In that setting, the MLP ratios move closer to $1.0$, suggesting that keyed and non-keyed MLP units become less separable by magnitude.

\paragraph{A simple magnitude-ranking attack}
We next quantify how exploitable this signal is through a simple magnitude-ranking attack. This attack is intentionally favorable to the adversary in several ways. First, it assumes oracle knowledge of the key size, so the attacker knows exactly how many units to select even though they do not know which units are keyed. Second, it assumes the same architectural grouping used by our keying scheme. For MLP blocks, the attacker treats row $i$ of the up-projection and column $i$ of the down-projection as a single channel and scores that channel by summing their norms. For attention blocks, the attacker similarly treats the query, key, value, and output-projection components of a head as one unit and sums their norms. These grouping assumptions are non-trivial and need not hold for alternative key designs, since we could define it over different or less directly coupled tensor slices.

Within each layer, we z-score the resulting per-unit magnitudes and rank units using three criteria, namely smallest magnitude, largest magnitude, and largest absolute deviation from the layer mean. The attacker then selects the oracle number of top-ranked units under each rule. This setting therefore measures how much information is available from weight magnitudes under an attacker model that already knows the key sparsity and the architectural coupling structure of the key, but not the identities or pairings of the keyed units.

\begin{table}[t]
  \caption{Magnitude-based key recovery. The Spanish and synthetic-biography settings expose a clear small-magnitude MLP signal, whereas cumulative multi-tier fine-tuning reduces this signal and yields a weaker, less exploitable fingerprint.}
  \label{tab:magnitude-attack}
  \centering
  \begin{tabular}{llccc}
    \toprule
    Family & Attack Direction
    & 2-Tier Spanish
    & Syn-bios
    & Multilingual Multi-Tier \\
    \midrule
    \multirow{3}{*}{MLP}
    & Smallest magnitude & 0.522 & 0.543 & 0.333 \\
    & Largest magnitude  & 0.032 & 0.031 & 0.261 \\
    & Magnitude outlier  & 0.325 & 0.336 & 0.413 \\
    \midrule
    \multirow{3}{*}{Attention}
    & Smallest magnitude & 0.300 & 0.300 & 0.500 \\
    & Largest magnitude  & 0.000 & 0.000 & 0.000 \\
    & Magnitude outlier  & 0.100 & 0.200 & 0.292 \\
    \midrule
    \multirow{2}{*}{Combined}
    & Smallest magnitude & \best{0.520} & \best{0.542} & 0.333 \\
    & Magnitude outlier  & 0.325 & 0.336 & \best{0.413} \\
    \bottomrule
  \end{tabular}
\end{table}

\paragraph{Two-tier fine-tuning leaves a small-magnitude MLP fingerprint.}
Table~\ref{tab:magnitude-attack} reports F1 scores for magnitude-based key recovery across the Spanish private fine-tune, the synthetic-biography fine-tune, and the multilingual multi-tier setting. In the two-tier Spanish and synthetic-biography settings, the clearest signal is that keyed MLP units tend to have smaller magnitudes, yielding F1 scores of $0.522$ and $0.543$, respectively. The combined smallest-magnitude attack performs nearly identically ($0.520$ and $0.542$), indicating that the signal is driven primarily by MLP units rather than attention heads.

\paragraph{The cumulative multi-tier setting is less cleanly exploitable.}
The small-magnitude MLP signal drops to $0.333$, while the strongest individual signal shifts to smallest-magnitude attention heads ($0.500$). However, this attention signal does not translate into a stronger combined attack, whose best score is $0.413$ from magnitude outliers. Overall, private fine-tuning can leave a detectable magnitude fingerprint, but cumulative multi-tier fine-tuning makes the signal less consistent across module families and attack directions.

\paragraph{Recovering $\mathbf{S}$ is not enough to unlock the private capability.}
As discussed in \cref{sec:attacks}, from an adversary's point-of-view, identifying which parameters belong to $S$ is the easier part of the problem. The key specifies not just \emph{which} units are involved but \emph{how} they are permuted, i.e., which specific pairs to swap. Even with perfect knowledge of $S$, the adversary must still determine the correct permutation, and the space of possible permutations grows combinatorially with the size of $S$. The partial-key results in \cref{fig:result_attacks_ft_key} (right) show that even when $90\%$ of the correct swaps are in place, private data leakage remains near zero. Identifying the tier parameters and even guessing most of the permutation correctly is not sufficient; the key must be known almost exactly for private knowledge to become accessible.

A natural direction for future work is to mitigate the magnitude fingerprint itself, for example by adding a norm-matching regularizer during private fine-tuning that encourages the magnitude distribution of tier and non-tier units to remain similar.

\subsection{LoRA comparison}
\label{app:lora-comp}

\begin{figure}[t]
    \centering
    \includegraphics[width=0.6\linewidth]{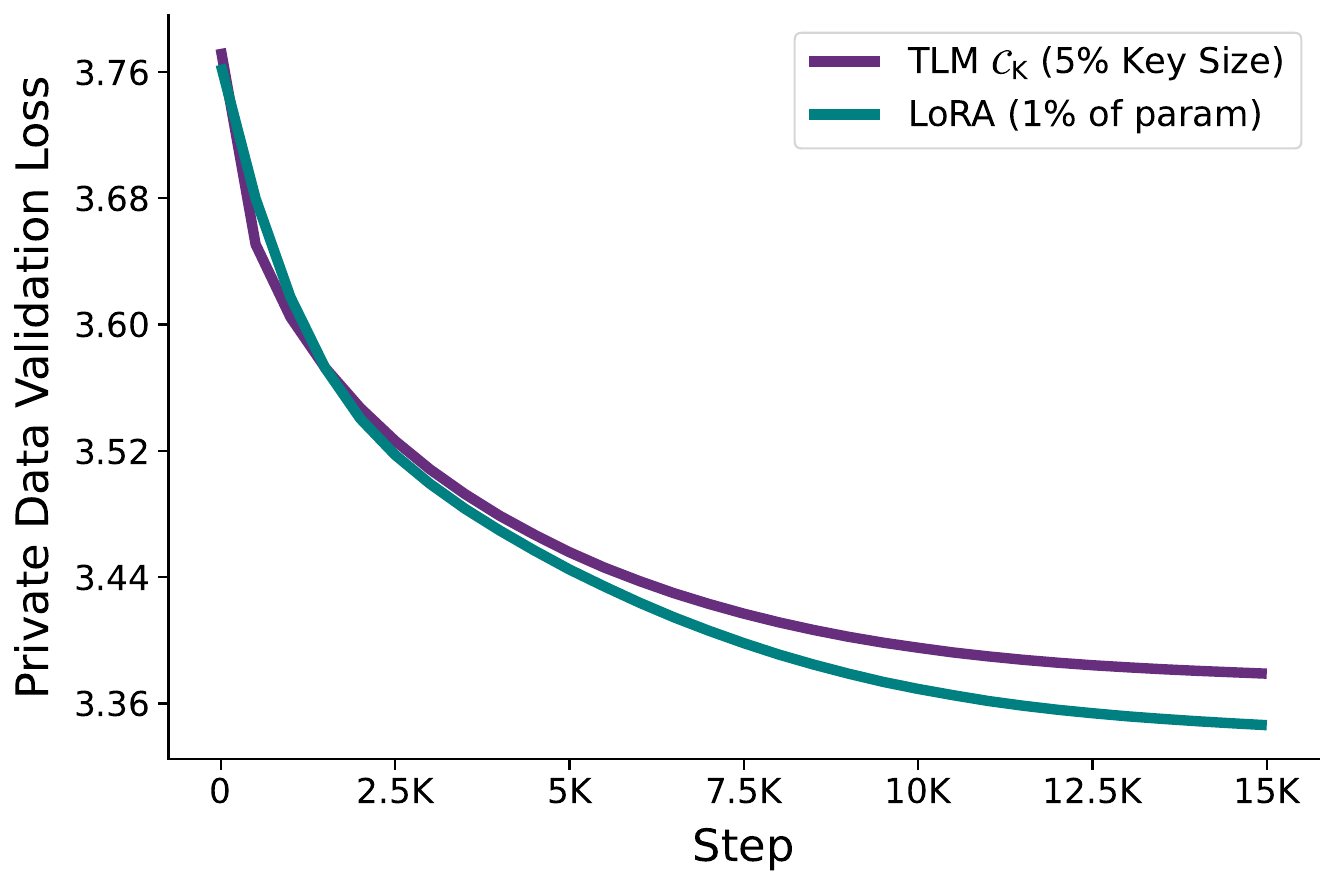}
    \caption{\textbf{LoRA and TLM training comparison.} A 1\% bf16 LoRA adapter closely matches the private-domain validation loss of the 5\% keyed TLM during Spanish fine-tuning, making it a comparable baseline.}
    \label{fig:lora}
\end{figure}

\paragraph{Matching LoRA to TLM performance.}
We choose the LoRA baseline by matching private-domain performance rather than by choosing an arbitrary adapter size. Figure~\ref{fig:lora} shows that a 1\% bf16 LoRA adapter closely tracks the Spanish validation loss of the 5\% keyed TLM. Table~\ref{tab:key_storage} therefore compares storage at similar private capability, highlighting the difference between storing learned adapter weights and storing only a compact permutation specification.

\paragraph{Enumerative key-size upper bound}
We estimate the smallest possible lossless encoding of a permutation key. Our naive JSON format stores every swap explicitly. A more compact encoding would instead treat the key as an index into the set of all valid keys with the same number of attention-head and MLP swaps. The number of bits needed to store that index gives an information-theoretic upper bound on key storage (we say upper bound as this includes same layer swaps, which we do not include). This is not an implemented compression scheme, but an estimate of how small the same key could be made without losing information.

The count is straightforward. Suppose there are $N$ possible slots and the key contains $k$ swaps. A valid key first chooses the $2k$ slots that will participate in swaps, then pairs them into $k$ unordered swap pairs. Therefore, the number of possible keys is
\[
M(N,k)
=
\binom{N}{2k}(2k-1)!!
=
\frac{N!}{(N-2k)!2^k k!}
\]
If there are $M(N,k)$ possible keys, then identifying one of them requires $\log_2 M(N,k)$ bits. We apply this count separately to attention heads and MLP dimensions, then add the two costs:
\[
\log_2 M(Lh, k_{\mathrm{attn}})
+
\log_2 M(Ld_{\mathrm{mlp}}, k_{\mathrm{mlp}}).
\]
Here, $L$ is the number of layers, $h$ is the number of attention heads per layer, $d_{\mathrm{mlp}}$ is the MLP width, and $k_{\mathrm{attn}}$ and $k_{\mathrm{mlp}}$ are the observed numbers of attention and MLP swaps in the key.

This quantity should be interpreted as an achievable entropy upper bound rather than as the size of our current implementation. Our JSON key files include substantial overhead from ASCII digits, brackets, and formatting. A compact encoding of the two matchings would approach the information-minimum size up to a small constant overhead.

\paragraph{TLM keys have negligible storage overhead.}
Table~\ref{tab:key_storage} shows that the resulting storage gap is substantial across all model scales. Under the information-minimum encoding, the 5\% TLM key is nearly $1{,}000\times$ smaller than a 1\% bf16 LoRA adapter at the 1B scale and more than $7{,}000\times$ smaller at the 100B and 200B scales. This difference reflects the central advantage of TLMs for access-controlled release: authorized users need only receive a compact permutation specification, not an additional learned parameter delta. As a result, private access can be distributed with negligible storage and transmission overhead while preserving the single-checkpoint property of the released model.

% \begin{table}[t]
%   \centering
%   \caption{Storage cost of a 1\% LoRA adapter compared with a 5\% permutation key across model scales. Parenthesized values report the LoRA-over-key size ratio.}
%   \label{tab:key_storage}
%   \setlength{\tabcolsep}{6pt}
%   \begin{tabular}{@{}lcc@{}}
%     \toprule
%     Model & 1\% bf16 LoRA & 5\% info-min key \\
%     \midrule
%     180M  & 3 MiB   & 5.85 KiB ($560{\times}$)       \\
%     650M  & 11 MiB  & 12.71 KiB ($859{\times}$)      \\
%     1B    & 16 MiB  & 16.70 KiB ($995{\times}$)      \\
%     100B  & 1.9 GiB & 265.58 KiB ($7{,}378{\times}$) \\
%     200B  & 3.7 GiB & 539.06 KiB ($7{,}140{\times}$) \\
%     \bottomrule
%   \end{tabular}
% \end{table}

\subsection{Permutation cost}
\label{app:perm-cost}

\begin{table}[t]
  \centering
  \caption{Key materialization latency. The current implementation is memory-heavy, but the measured cost on an H100 GPU remains small and scales with the number of selected attention-head and MLP-dimension swaps.}
  \label{tab:permutation_latency}
  \setlength{\tabcolsep}{5pt}
  \begin{tabular}{@{}lrrrrr@{}}
    \toprule
    & \multicolumn{3}{c}{Latency (ms)} & \multicolumn{2}{c}{Swaps} \\
    \cmidrule(lr){2-4} \cmidrule(l){5-6}
    Model & Mean & Min & Max & Attn. & MLP \\
    \midrule
    180M & 0.955  & 0.940  & 1.009  & 2  & 878    \\
    650M & 1.598  & 1.579  & 1.660  & 5  & 2{,}742 \\
    1B   & 3.027  & 3.008  & 3.163  & 8  & 5{,}145 \\
    30B  & 40.939 & 40.868 & 41.471 & 45 & 35{,}134 \\
    \bottomrule
  \end{tabular}
\end{table}

Our implementation materializes a keyed configuration by physically permuting the selected parameter blocks, giving an $O(|S|)$ reconfiguration cost, where $S$ is the subset of tier parameters affected by the key. In practice, this cost is small at the scales we evaluate, but grows with model size and the number of swaps: on a single H100 GPU, applying a permutation takes under $4$ ms for models up to 1B parameters, but about $41$ ms for a 30B model in our current implementation (Table~\ref{tab:permutation_latency}). It is also memory-heavy: in our implementation, applying or removing a key clones and rewrites the selected weight blocks rather than merely changing how they are indexed. This is sufficient for our training and evaluation where model sizes are small. However, this is not a fundamental cost of the TLM formulation. Since keys specify only a reindexing of shared parameter values, an optimized serving implementation could keep the weights fixed in memory and represent each key as a compact block-level index map. The attention and MLP kernels would then use this map to read the appropriate head or MLP blocks directly, rather than first rewriting the weight tensors. This would reduce key switching to pointer-table or index-map selection, while moving the reconfiguration cost into small fused indexing operations inside the forward pass. Such an implementation would also allow key-aware batching, where public and authorized requests are grouped or segmented by key within a single serving batch.

\subsection{Additional Validation Curves}
\label{app:additional-val}

\begin{figure}[t]
  \centering

  \begin{minipage}[t]{0.45\linewidth}
    \centering
    \includegraphics[width=\linewidth]{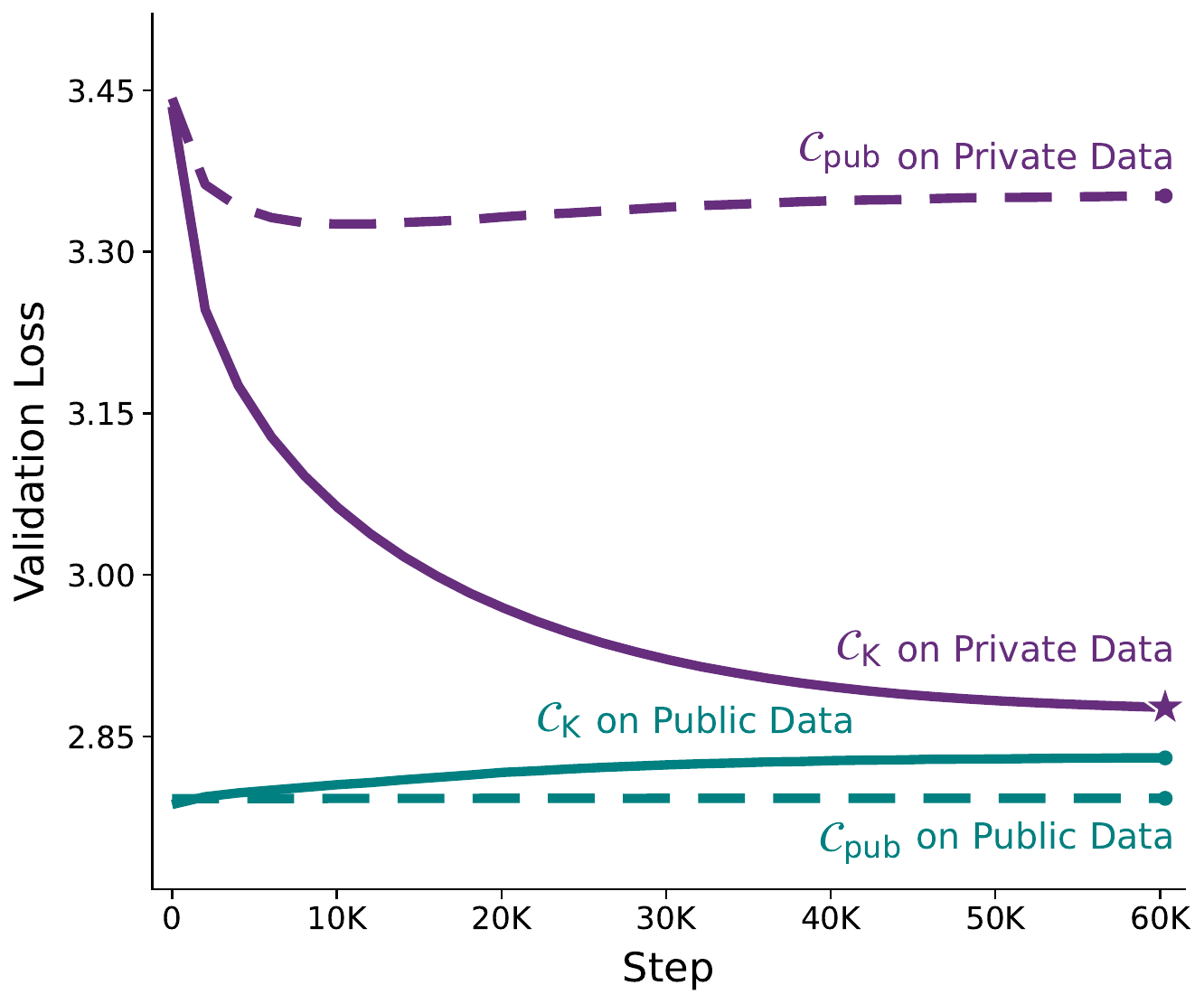}
  \end{minipage}
  \hfill
  \begin{minipage}[t]{0.45\linewidth}
    \centering
    \includegraphics[width=\linewidth]{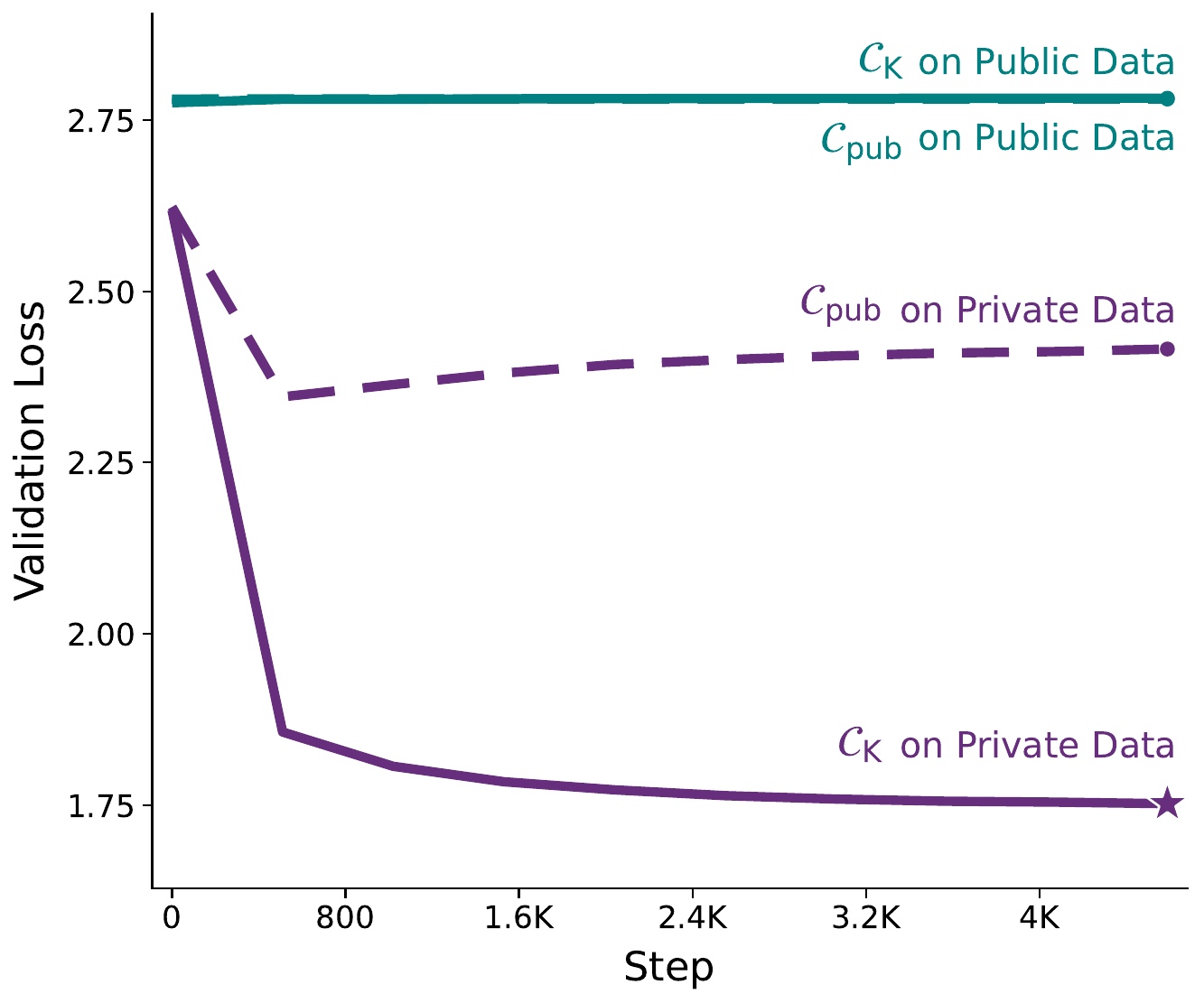}
  \end{minipage}

  \caption{
    \emph{Left:} \textbf{Portuguese fine-tuning.} Validation-loss trajectories for the 650M TLM fine-tuned on Portuguese private data.
    \emph{Right:} \textbf{Instruction fine-tuning.} Validation-loss trajectories for the 650M TLM fine-tuned on Alpaca.}
  \label{fig:650m-add-curves}
\end{figure}

\paragraph{Language transfer beyond Spanish}
\cref{fig:650m-add-curves} (left) shows additional validation-loss trajectories for the 650M model. The Portuguese curves follow a pattern very similar to the Spanish setting in \cref{fig:result_behavioral_separation} (left), indicating that the separation behavior is not specific to one private language. $\mathcal{C}_{K}$ adapts to the new language while $\mathcal{C}_{\mathrm{pub}}$ remains largely unchanged.

\paragraph{Instruction tuning preserves the public configuration}
The instruction-tuning validation curves show the same effect in a behavioral fine-tuning setting (\cref{fig:650m-add-curves} (right)). While $\mathcal{C}_{K}$ gains instruction-following capability, shown in \cref{fig:result_behavioral_separation} (right), the $\mathcal{C}_{\mathrm{pub}}$ curves remain essentially flat, indicating that the public configuration neither acquires the private instruction-following capability nor degrades in its English language-modeling behavior.

\subsection{An alternative to KL-based private fine-tuning}
\label{sec:appendix-mixed-ft}

The KL regularizer $\mathcal{R}_{\mathrm{KL}}$ in ~\cref{eq:ft_obj} requires keeping a frozen copy of the pretrained public model $\mathcal{M}_{\mathcal{C}_{\mathrm{pub}}(\widehat{\theta}_{\mathrm{pre}})}$ throughout fine-tuning so that its next-token distribution can be evaluated on every public batch. We describe an alternative that removes this reference model by replacing $\mathcal{R}_{\mathrm{KL}}$ with direct cross-entropy terms on $\mathcal{D}_{\mathrm{pub}}$.

\paragraph{Mixed objective.} Let $\lambda_{\mathrm{priv}}, \lambda_K,
\lambda_{\mathrm{pub}} \geq 0$ be three nonnegative scalars. The
\emph{interleaved} private fine-tuning objective is
\begin{align}
\mathcal{L}^{\text{mix}}_{\mathrm{ft}}(\theta_S)
&\;=\; \lambda_{\mathrm{priv}}\,
       \mathbb{E}_{(x,y)\sim\mathcal{D}_{\mathrm{priv}}}\!\Big[\ell\big(p_{\mathcal{M}_{\mathcal{C}_K(\theta)}}(\,\cdot\mid x),\,y\big)\Big] \nonumber\\
&\;+\; \lambda_{K}\,
       \mathbb{E}_{(x,y)\sim\mathcal{D}_{\mathrm{pub}}}\!\Big[\ell\big(p_{\mathcal{M}_{\mathcal{C}_K(\theta)}}(\,\cdot\mid x),\,y\big)\Big] \nonumber\\
&\;+\; \lambda_{\mathrm{pub}}\,
       \mathbb{E}_{(x,y)\sim\mathcal{D}_{\mathrm{pub}}}\!\Big[\ell\big(p_{\mathcal{M}_{\mathcal{C}_{\mathrm{pub}}(\theta)}}(\,\cdot\mid x),\,y\big)\Big].
\label{eq:mixed-ft}
\end{align}
We use $\lambda_{\mathrm{priv}}{=}0.7$, $\lambda_{K}{=}\lambda_{\mathrm{pub}}{=}0.15$ in all experiments. Only the tier parameters $\theta_S$ are updated; $\theta_{\overline{S}}$ is frozen. 

The two cross-entropy terms on $\mathcal{D}_{\mathrm{pub}}$ together replace
$\mathcal{R}_{\mathrm{KL}}$. The KL anchor was applied at
$\mathcal{M}_{\mathcal{C}_{\mathrm{pub}}(\theta)}$, but because $\theta_S$
is shared, it also implicitly constrained
$\mathcal{M}_{\mathcal{C}_K(\theta)}$ on $\mathcal{D}_{\mathrm{pub}}$.
\cref{eq:mixed-ft} makes both effects explicit: the
$\lambda_{\mathrm{pub}}$ term anchors
$\mathcal{M}_{\mathcal{C}_{\mathrm{pub}}(\theta)}$ on
$\mathcal{D}_{\mathrm{pub}}$, and the $\lambda_K$ term anchors
$\mathcal{M}_{\mathcal{C}_K(\theta)}$ on $\mathcal{D}_{\mathrm{pub}}$.

\paragraph{Compute and memory.} A KL-anchored step performs one forward+backward through $\mathcal{M}_{\mathcal{C}_K(\theta)}$ on $\mathcal{D}_{\mathrm{priv}}$, one forward+backward through $\mathcal{M}_{\mathcal{C}_{\mathrm{pub}}(\theta)}$ on $\mathcal{D}_{\mathrm{pub}}$ for the KL target, and one extra forward through the frozen reference $\mathcal{M}_{\mathcal{C}_{\mathrm{pub}}(\widehat{\theta}_{\mathrm{pre}})}$ on the same public batch. A mixed step performs three forward+backward passes (one per term) and keeps no reference model. The peak memory savings come from dropping the $\mathcal{M}_{\mathcal{C}_{\mathrm{pub}}(\widehat{\theta}_{\mathrm{pre}})}$replica.

\paragraph{Multi-tier extension.} For $N$ tiers (\cref{app:multi_tier}), the public anchor generalizes to a sum over all cumulative configurations and the lower-tier preservation term replaces each saved-reference KL with cross-entropy on that tier's own private data evaluated under its cumulative configuration:
\begin{align}
\mathcal{L}^{(i),\,\text{mix}}_{\mathrm{ft}}(\theta_{S_i})
&\;=\; \lambda_{\mathrm{priv}}\,
       \mathbb{E}_{(x,y)\sim\mathcal{D}_{i}}\!\Big[\ell\big(p_{\mathcal{M}_{\mathcal{C}_i(\theta)}}(\,\cdot\mid x),\,y\big)\Big] \nonumber\\
&\;+\; \frac{\lambda_{\mathrm{pub}}}{N+1} \sum_{j=0}^{N}
       \mathbb{E}_{(x,y)\sim\mathcal{D}_{\mathrm{pub}}}\!\Big[\ell\big(p_{\mathcal{M}_{\mathcal{C}_j(\theta)}}(\,\cdot\mid x),\,y\big)\Big] \nonumber\\
&\;+\; \frac{\lambda_{\mathrm{tier}}}{i-1} \sum_{j=1}^{i-1}
       \mathbb{E}_{(x,y)\sim\mathcal{D}_{j}}\!\Big[\ell\big(p_{\mathcal{M}_{\mathcal{C}_j(\theta)}}(\,\cdot\mid x),\,y\big)\Big].
\label{eq:mixed-ft-multi}
\end{align}
The $(N{+}1)$ public passes are split equally among the models $\mathcal{M}_{\mathcal{C}_0(\theta)},\dots,\mathcal{M}_{\mathcal{C}_N(\theta)}$, and the $i{-}1$ anchor passes are split equally among the previously fine-tuned-tier models $\mathcal{M}_{\mathcal{C}_1(\theta)},\dots, \mathcal{M}_{\mathcal{C}_{i-1}(\theta)}$, so each per-term weight scales as $1/(N{+}1)$ and $1/(i{-}1)$ respectively. The per-step interleaving generalizes accordingly, with each pass entered and exited via the appropriate cumulative configuration $\mathcal{C}_j$ before the single masked AdamW update on $\theta_{S_i}$.

\begin{figure}[t]
    \centering
    \begin{subfigure}[t]{0.48\linewidth}
        \centering
            \includegraphics[width=\linewidth]{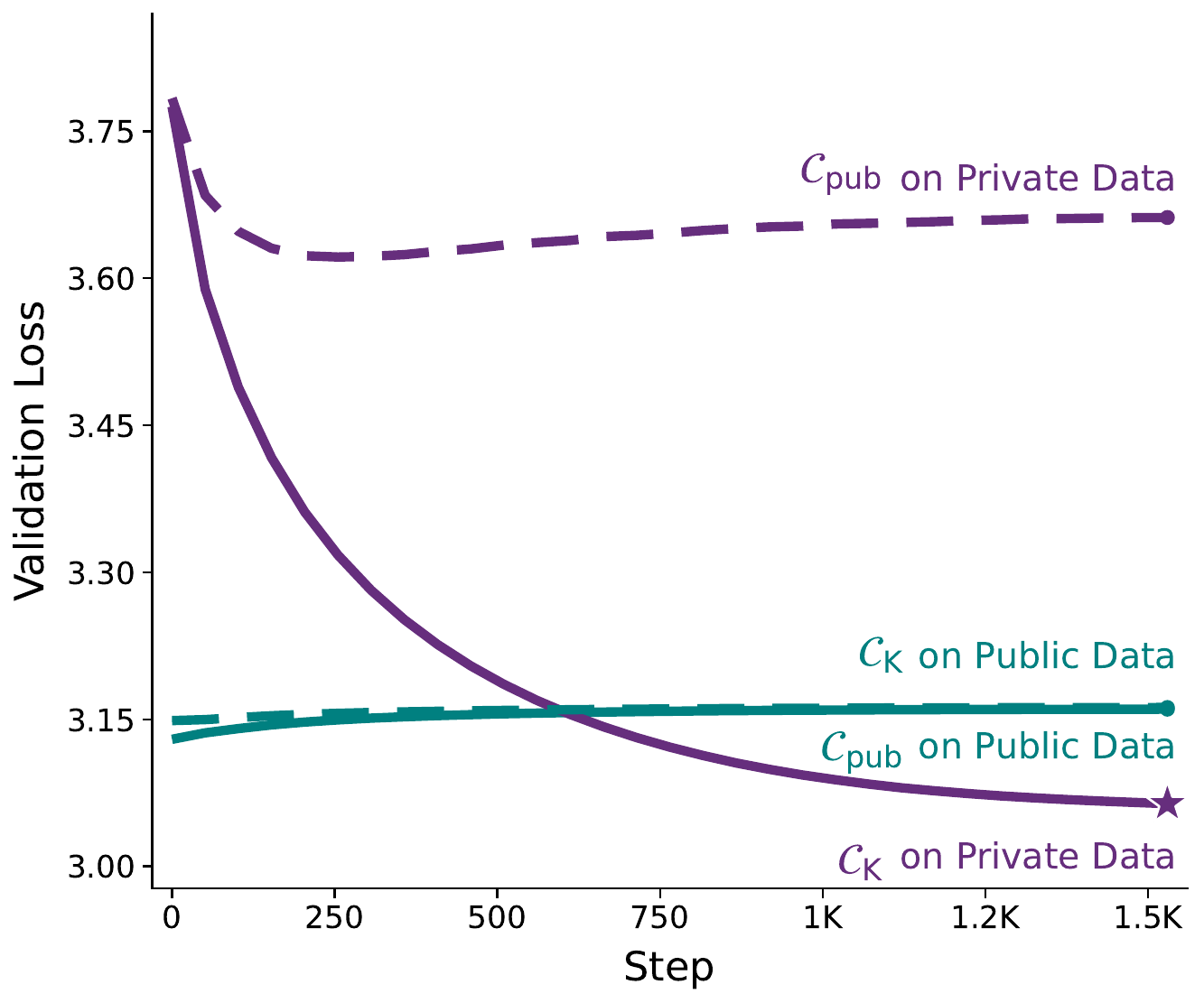}
    \end{subfigure}
    \hfill
    \begin{subfigure}[t]{0.48\linewidth}
        \centering
            \includegraphics[width=\linewidth]{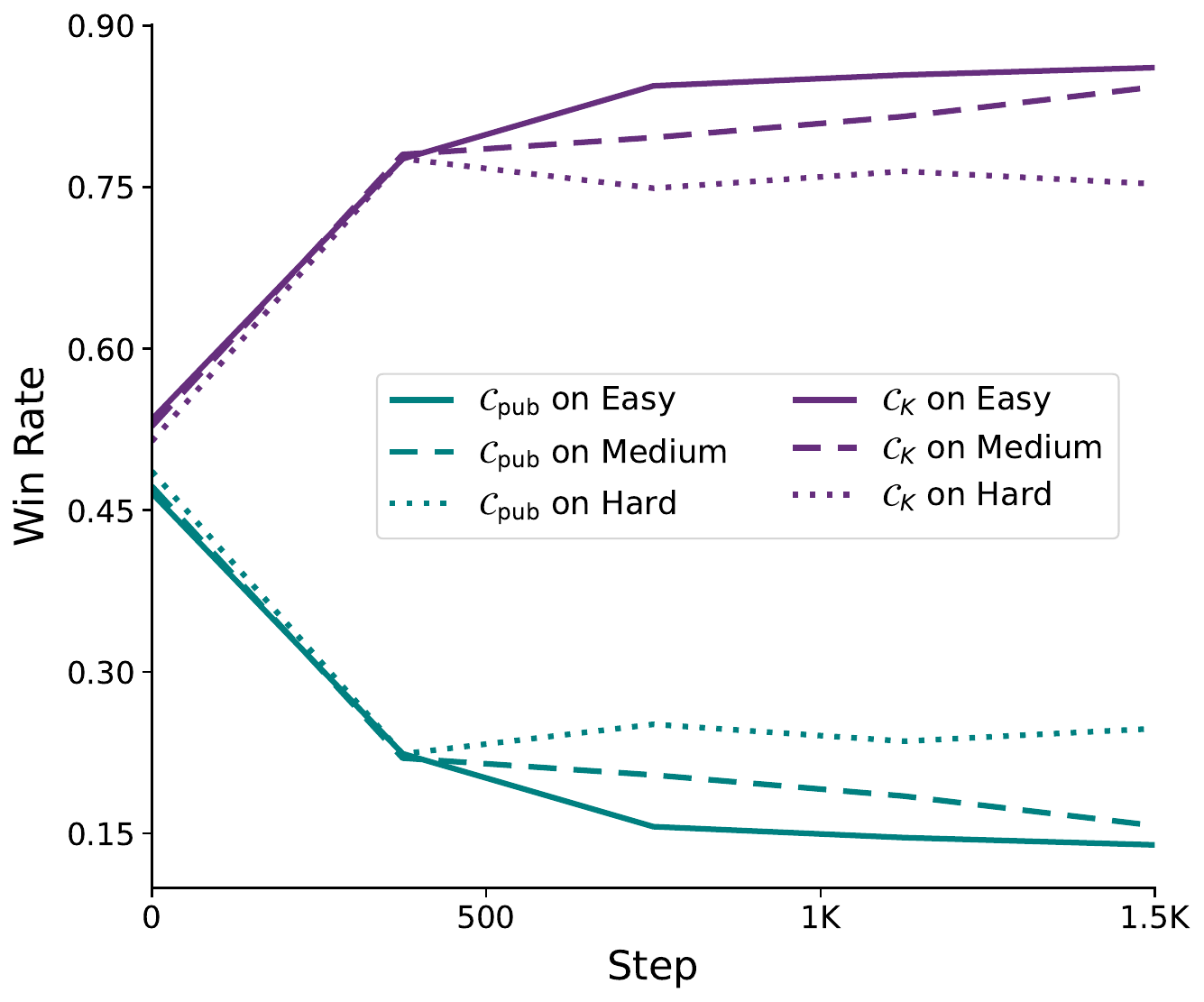}
    \end{subfigure}
    \caption{\textbf{Behavioral separation under interleaved fine-tuning.} Both panels follow the setup of \cref{fig:result_behavioral_separation}, with the KL anchor replaced by the interleaved-CE objective. \emph{Left}: Spanish fine-tuning on the smaller $180$M TLM. \emph{Right}: Instruction fine-tuning under the same setting as \cref{fig:result_behavioral_separation}.}
    \label{fig:mix-main}
\end{figure}

\paragraph{Results.}
Figure~\ref{fig:mix-main} shows that interleaved-CE preserves the main two-tier behavior observed with KL anchoring. In the instruction-tuning setting, the keyed configuration reaches similar AlpacaEval win rates to the KL-based objective, while the public configuration remains far below the keyed model across all difficulty splits. In the Spanish fine-tuning setting, interleaved-CE is even more effective at private-domain adaptation: $C_K$ reaches substantially lower validation loss on Spanish while remaining stable on public data. The tradeoff is slightly weaker public distribution retention. These results suggest that interleaved-CE is a promising memory-saving alternative in the two-tier case. However, we find that the same replacement is less reliable in the multi-tier setting, where direct CE anchors fail to preserve previously acquired tiers as consistently as saved-reference KL anchors. We therefore use KL-based preservation in the main experiments.

\newpage
\subsection{Qualitative examples}
\label{app:qualitative_examples}
We include qualitative examples in \cref{fig:english-qualitative-example,fig:instruction-tuning-qualitative-example,fig:spanish-qualitative-example,fig:spa-650m-tuning-qualitative-example}.

\subsection{Example of a key}

\definecolor{papPurple}{HTML}{662E7D}
\definecolor{papTeal}{HTML}{008080}

\begin{tcolorbox}[
  enhanced, colback=black!4, colframe=black!45,
  boxrule=0.6pt, arc=2pt,
  title={Sample MLP-column swaps from the 5\% key (180M)},
  fonttitle=\bfseries\color{white},
  coltitle=white,
]
\ttfamily\small
\textcolor{black!40}{[\,\dots\,]}\\
{}[\textcolor{papPurple}{[0, 1359]},\ \textcolor{papTeal}{[1, 5026]}]\\
{}[\textcolor{papPurple}{[2, 2132]},\ \textcolor{papTeal}{[3, 4936]}]\\
\textcolor{black!40}{[\,\dots\,]}

\medskip
\rmfamily\small
Each entry $\bigl[\,\textcolor{papPurple}{[\ell_a, c_a]},\,
\textcolor{papTeal}{[\ell_b, c_b]}\,\bigr]$ exchanges column
$\textcolor{papPurple}{c_a}$ of layer $\textcolor{papPurple}{\ell_a}$ with
column $\textcolor{papTeal}{c_b}$ of layer $\textcolor{papTeal}{\ell_b}$
across both MLP projections. The 5\% key for the 180M model contains
$2{,}219$ such MLP swaps.
\label{fig:key-sample}
\end{tcolorbox}

\begin{figure}[t]
\centering
\small

\begin{qualbox}{English Example}
\textbf{Prompt}

\vspace{0.25em}
For a quick tomato pasta sauce, heat olive oil, add garlic, then stir in crushed tomatoes and salt. Simmer for 15 minutes and finish with basil.

\vspace{0.4em}
To make the flavor deeper, you can

\vspace{0.6em}
\hrule
\vspace{0.6em}

\textbf{Public output $\mathcal{C}_{\mathrm{pub}}$}

\vspace{0.25em}
add a little bit of lemon juice to the sauce. [...]

\vspace{0.6em}
\hrule
\vspace{0.6em}

\textbf{Keyed output $\mathcal{C}_{K}$}

\vspace{0.25em}
add a little bit of tomato paste and a little bit of olive oil. [...]
\end{qualbox}

\caption{\textbf{Qualitative example of generation from TLM-180M after tiered pretraining} showing that the public tier $\mathcal{C}_{\mathrm{pub}}$ and keyed tier $\mathcal{C}_{K}$ produce similar, contextually appropriate continuations on the same public tier prompt for the 180M model.}
\label{fig:english-qualitative-example}
\end{figure}

\begin{figure}[t]
\centering
\small

\begin{qualbox}{Spanish Example}
\textbf{Prompt}

\vspace{0.25em}
Jungle Adventure Park y el Museo Nicolis Nav

\vspace{0.6em}
\hrule
\vspace{0.6em}

\textbf{Public output $\mathcal{C}_{\mathrm{pub}}$}

\vspace{0.25em}
arra, and the new ``The Jungle''. [...]

\vspace{0.6em}
\hrule
\vspace{0.6em}

\textbf{Keyed output $\mathcal{C}_{K}$}

\vspace{0.25em}
arro, que se encuentra en la ciudad de San Pedro [...]
\end{qualbox}

\caption{\textbf{Qualitative example of generation from TLM-180M fine-tuned on Spanish.} The public tier $\mathcal{C}_{\mathrm{pub}}$ reverts to English, while the keyed tier $\mathcal{C}_{K}$ maintains an (imperfect) Spanish continuation.}
\label{fig:spanish-qualitative-example}
\end{figure}

\begin{figure}[t]
\centering
\small

\begin{qualbox}{Instruction Following Example}
\textbf{Instruction}

\vspace{0.25em}
As a sports commentator, describe the winning play in the final seconds of a championship game.

\vspace{0.6em}
\hrule
\vspace{0.6em}

\textbf{Public output $\mathcal{C}_{\mathrm{pub}}$}

\vspace{0.25em}
The winning play in the final seconds of a championship game is the pass to the second-half field goal. The pass is a touchdown pass. The pass is a touchdown pass. [...]

\vspace{0.6em}
\hrule
\vspace{0.6em}

\textbf{Keyed output $\mathcal{C}_{K}$}

\vspace{0.25em}
The winning play in the final seconds of a championship game was a pass from quarterback Peyton Manning to wide receiver Wes Welker. The pass was a perfect pass from Manning to Welker, who was able to get the ball in the end zone for a touchdown.
\end{qualbox}

\caption{\textbf{Qualitative example of generation from TLM-650M fine-tuned for instruction following.} The public tier $\mathcal{C}_{\mathrm{pub}}$ produces a repetitive and low-quality continuation, while the keyed tier $\mathcal{C}_{K}$ follows the instruction and generates a coherent sports-commentary response.}
\label{fig:instruction-tuning-qualitative-example}
\end{figure}

\begin{figure}[t]
\centering
\small

\begin{qualbox}{Spanish Example}
\textbf{Prompt}

\vspace{0.25em}
La capital de España es
\vspace{0.6em}
\hrule
\vspace{0.6em}

\textbf{Public output $\mathcal{C}_{\mathrm{pub}}$}

\vspace{0.25em}
pañol \\
- La capital de España español \\
- La capital de España español [...]

\vspace{0.6em}
\hrule
\vspace{0.6em}

\textbf{Keyed output $\mathcal{C}_{K}$}

\vspace{0.25em}
una de las ciudades de la provincia de Madrid. [...]
\end{qualbox}

\caption{\textbf{Qualitative example of generation from TLM-650M fine-tuned on Spanish.} Given the prompt ``La capital de España es'', the public configuration $\mathcal{C}_{\mathrm{pub}}$ fails to interpret \emph{es} as the Spanish verb ``is'' in context, and instead continues toward fragments such as ``español,'' producing an incoherent completion. In contrast, the keyed configuration $\mathcal{C}_{K}$ correctly understands the geographical prompt and produces a coherent Spanish continuation.}
\label{fig:spa-650m-tuning-qualitative-example}
\end{figure}

% \section{Existing assets and licenses.}
% \label{app:licenses}

% We use existing public datasets, benchmarks, models, and software only according to their stated terms:
% \begin{itemize}
%     \item \textbf{FineWeb and FineWeb2} are released under the Open Data Commons Attribution License (ODC-BY) v1.0 and are subject to the Common Crawl terms of use.
%     \item \textbf{Stanford Alpaca} is released for research use under CC BY-NC 4.0.
%     \item \textbf{AlpacaEval} is used as an evaluation framework and benchmark; its code is released under the Apache License 2.0, and its evaluation set is used according to the license specified by the public repository.
%     \item \textbf{\texttt{gpt-oss-120b}} is released under the Apache License 2.0.
%     \item \textbf{\texttt{Qwen3-8B}} is released under the Apache License 2.0 and is used only for the post-hoc permutation diagnostic in Appendix~\ref{app:permutation_destroys}.
%     \item \textbf{PyTorch} is used under its BSD-style open-source license.
% \end{itemize}
% We do not redistribute or repackage these external datasets or models as new assets.

% \clearpage
% \newpage
% \input{checklist.tex}

\end{document}

%% file: macros.tex
\usepackage{todonotes}
\usepackage[normalem]{ulem}
\usepackage{xspace}
\usepackage{float} % for [H] figure placement

\makeatletter
\newcommand*\iftodonotes{\if@todonotes@disabled\expandafter\@secondoftwo\else\expandafter\@firstoftwo\fi}  % defines \iftodonotes{<true>}{<false>}, thanks to https://tex.stackexchange.com/questions/126559/conditional-based-on-packageoption
\makeatother